%% file: sgx.tex
\PassOptionsToPackage{hyphens}{url}
\PassOptionsToPackage{table}{xcolor}
\documentclass[sigconf, nonacm]{acmart}

\pdfoutput=1

\input{header}

\input{graphs}

\begin{document}

\title{Dissecting BFT Consensus: In Trusted Components we Trust!}

\author{Suyash Gupta$^{\dagger *}$, Sajjad Rahnama$^*$, Shubham Pandey$^*$, Natacha Crooks$^{\dagger}$, Mohammad Sadoghi$^*$}
\affiliation{%
\institution{ 
\setlength{\tabcolsep}{5mm}
\begin{tabular}{cc}
   $^{\dagger}$SkyLab &  $^*$Exploratory Systems Lab\\
   University of California, Berkeley}	& University of California, Davis \\
\end{tabular}
}

\begin{abstract}
\input{abstract}
\end{abstract}

\maketitle

\input{intro}

\input{prelim}

\input{back}

\input{trusted}

\input{liveness}
\input{unsafe}

\input{parallel}

\input{extend}

\input{proofs}
\input{eval}

\input{related}

\input{concl}

\bibliographystyle{ACM-Reference-Format}
\bibliography{refined}

\appendix
\input{flexibft}

\end{document}

%% file: header.tex
\usepackage{amsthm,amsmath,amsfonts}
\usepackage[inline]{enumitem}
\usepackage{caption,subcaption}
\usepackage{balance}
\usepackage[binary-units]{siunitx}
\DeclareSIUnit{\nothing}{\relax}
\usepackage{graphicx}
\usepackage{pifont}
\usepackage{array,booktabs}
\usepackage{multirow}
\usepackage{color}
\usepackage[normalem]{ulem} % Defines strikeout Breaks bibtex

\newcommand{\revise}[1]{#1}

\usepackage{tikz,pgfplots,pgfplotstable}
\usetikzlibrary{decorations.pathreplacing}

\newtheorem{theorem}{Theorem}
\newtheorem{definition}[theorem]{Definition}

\newcommand{\changebars}[2]{#1}

\newcommand{\cmark}{\ding{51}}%
\newcommand{\xmark}{\ding{55}}%

\newcolumntype{C}[1]{>{\centering\let\newline\\\arraybackslash\hspace{0pt}}m{#1}}

\newcommand\HLine[1]{\noalign{\hrule height #1}}

\newlength{\Oldarrayrulewidth}

\newcommand{\Replicas}{\mathcal{R}}
\newcommand{\Service}{\mathcal{S}}
\newcommand{\Clients}{\mathcal{C}}

\newcommand{\Faulty}{\mathcal{F}}
\newcommand{\n}{\mathbf{n}}
\newcommand{\f}{\mathbf{f}}

\newcommand{\Replica}[1][r]{\textsc{#1}}
\newcommand{\Primary}[1][p]{\textsc{#1}}
\newcommand{\Client}[1][c]{\MakeLowercase{#1}}
\newcommand{\Trust}[1]{\textsc{t}_{#1}}

\newcommand{\Name}[1]{\textsc{#1}}
\newcommand{\PName}[1]{\textsc{#1}}
\newcommand{\RSM}{\PName{Rsm}}
\newcommand{\BFT}{\PName{bft}}
\newcommand{\CFT}{\PName{cft}}
\newcommand{\trustBFT}{\PName{trust-bft}}
\newcommand{\trustBFTb}{{\bf \PName{trust-bft}}}
\newcommand{\trustBFTc}{\PName{Opbft-ea}}
\newcommand{\lftBFT}{\PName{FlexiTrust}}
\newcommand{\ZZ}{\PName{Zyzzyva}}

\newcommand{\pbft}{\PName{Pbft}}
\newcommand{\pbftea}{\PName{Pbft-EA}}
\newcommand{\minLFT}{\PName{Flexi-BFT}}
\newcommand{\minLFTZ}{\PName{Flexi-ZZ}}
\newcommand{\hotstuff}{\PName{HotStuff}}
\newcommand{\hotstuffm}{\PName{HotStuff-M}}

\newcommand{\MinBFT}{\PName{MinBFT}}
\newcommand{\MinZZ}{\PName{MinZZ}}
\newcommand{\CheapBFT}{\PName{CheapBFT}}
\newcommand{\Trinc}{\PName{Trinc}}
\newcommand{\Hybster}{\PName{Hybster}}
\newcommand{\Damsys}{\PName{Damysus}}
\newcommand{\ResilientDB}{\PName{ResilientDB}}

\newcommand{\DS}{\Name{DS}}
\newcommand{\MAC}{\Name{MAC}}
\newcommand{\Col}[1]{\texttt{Col#1}}

\newcommand{\Transaction}[1][t]{\MakeUppercase{#1}}
\newcommand{\MName}[1]{\textsc{#1}}
\newcommand{\Message}[2]{\textsc{#1}(#2)}
\newcommand{\SignMessage}[2]{\langle#1\rangle_{#2}}
\newcommand{\Hash}[1]{\texttt{Hash}(#1)}
\newcommand{\foo}{{\tt foo()}}
\newcommand{\Digest}{\Delta}
\newcommand{\tid}{q}
\newcommand{\tct}{k}
\newcommand{\knew}{k_{new}}
\newcommand{\Sig}{\sigma}
\newcommand{\Result}{{\bf r}}

\newcommand{\abs}[1]{\lvert #1 \rvert}
\newcommand{\union}{\cup}

\newcommand{\difference}{\setminus}
\newcommand{\BigO}[1]{\mathcal{O}(#1)}

\DeclareSIUnit\k{k}
\DeclareSIUnit\ms{ms}
\DeclareSIUnit\GB{GiB}
\DeclareSIUnit\B{B}
\DeclareSIUnit{\txn}{txn}
\DeclareSIUnit{\batch}{batch}
\newtheorem{claim}{Claim}

\newcommand{\RN}[1]{%
  \textup{\expandafter{\romannumeral#1}}%
}

\usepackage[noend]{algorithmic}
\newenvironment{myprotocol}{
    \hrule
    \smallskip
    \scriptsize
    \algsetup{linenosize=\tiny}
    \begin{algorithmic}[1]
        \newcommand{\SPACE}{\item[]}	
	\newcommand{\GETS}{:=}
        \newcommand{\TITLE}[2]{\item[] \textbf{\underline{##1}} (##2) \textbf{:}\\[0.5pt]}
        \makeatletter
            \newcommand{\EVENT}[1]{\STATE \textbf{event} ##1 \textbf{do}\begin{ALC@g}}
            \newcommand{\ENDEVENT}{\end{ALC@g}}
        \makeatother
	
	\makeatletter
            \newcommand{\FUNC}[2]{\STATE \textbf{function} \textbf{##1} (##2) \begin{ALC@g}}
            \newcommand{\ENDFUNC}{\end{ALC@g}}
        \makeatother

	\newcommand{\INITIAL}[2]{\item[] \textbf{\underline{##1}} ##2\\[0.5pt]}
	
}{
    \end{algorithmic}
    \smallskip
    \hrule
}

%% file: graphs.tex
\usepackage{xcolor}
\usepackage{tikz,pgfplots,pgfplotstable}
\usetikzlibrary{arrows.meta}
\tikzset{
    >=Stealth,
    plot/.append style={baseline,scale=0.45},
    label/.append style={font=\strut\footnotesize},
    dot/.style={circle,scale=0.35,draw=black,fill=black},
}
\pgfplotscreateplotcyclelist{mycyclelist}{
    red   ,every mark/.append style={solid,fill=\pgfplotsmarklistfill},mark=*\\
    blue!90!white!90     ,every mark/.append style={solid,fill=\pgfplotsmarklistfill},mark=square*\\
    blue    ,every mark/.append style={solid,fill=\pgfplotsmarklistfill},mark=triangle*\\
    black    ,every mark/.append style={solid,fill=\pgfplotsmarklistfill},mark=pentagon*\\
    brown   ,every mark/.append style={solid,fill=\pgfplotsmarklistfill},mark=o\\
    orange  ,every mark/.append style={solid,fill=\pgfplotsmarklistfill},mark=halfcircle*\\
    teal  ,every mark/.append style={solid,fill=\pgfplotsmarklistfill,rotate=180},mark=halfdiamond*\\
    green  ,every mark/.append style={solid,fill=\pgfplotsmarklistfill,rotate=180},mark=diamond*\\
}
\pgfplotscreateplotcyclelist{mycyclelistex}{
    black   ,every mark/.append style={solid,fill=\pgfplotsmarklistfill},mark=*\\
    red     ,every mark/.append style={solid,fill=\pgfplotsmarklistfill},mark=square*\\
}
\pgfplotsset{
    tick label style={font=\Large},
    legend style={font=\large,cells={anchor=west},text=black},
    title style={font=\Large},
    label style={font=\normalsize},
    width=262.5pt,
    height=185pt,
    every axis/.append style={
            ylabel near ticks,
            xlabel near ticks,
            mark size=3pt,
            cycle list name=mycyclelist,
            font=\Large,
            y tick label style={
                    /pgf/number format/precision=1,
                    /pgf/number format/fixed,
                    /pgf/number format/fixed zerofill
                }
        },
}

\newcommand{\axisFail}{Number of Replicas (in $\f$)}
\newcommand{\axisRegions}{Number of Regions}
\newcommand{\axisTP}{Throughput (\si{\txn\per\second})}
\newcommand{\axisLat}{Latency (\si{\second})}
\newcommand{\axisbatches}{Batch size}

\pgfplotstableread{
 X     A2M     MinBFT  MinZZ   CT      FBFT    FZZ     PBFT    ZZ
 4     70303   103108  111884  74769   176217  205574  158962  225230
 8     60635   90090   90593   64144   147438  189264  112687  210455    
 16    49929   65130   67251   49750   109450  120653  72513   154631
 24    42042   50618   49075   42247   84843   97902   61021   98685
 32    37365   38768   41642   36703   71278   80857   40881   81772       
}\graphAtp

\pgfplotstableread{
X     A2M     MinBFT  MinZZ   CT      FBFT    FZZ       PBFT      ZZ
4     0.781   0.528   0.485   0.618   0.301   0.258     0.345     0.234
8     0.906   0.602   0.604   0.732   0.366   0.284     0.492     0.256  
16    1.114   0.858   0.819   0.96    0.503   0.472     0.781     0.355
24    1.342   1.107   1.126   1.142   0.659   0.586     0.935     0.562
32    1.529   1.46    1.335   1.318   0.793   0.695     1.421     0.687
}\graphAlt

\pgfplotstableread{
X   A2M     MinBFT  MinZZ   CT      FBFT    FZZ     PBFT    ZZ
1   0.51    0.268	0.205   0.618   0.18    0.178   0.156   0.085
2   0.776   0.556	0.52    0.732   0.504   0.399   0.503   0.367
3   0.991   0.889	0.798   1.114   0.721   0.715   0.8	    0.612
4   1.083   0.971	0.909   1.336   0.845   0.767   0.746   0.654
5   1.316   1.296	1.313   1.555   1.204   1.038   1.005   0.868
}\graphBlt

\pgfplotstableread{
X   A2M     MinBFT  MinZZ   CT      FBFT    FZZ     PBFT    ZZ
1   7676    14219   18369   7761    21226   22340   24158   42882
2   60735   85707   91003   64144   124711  158381  110946  171749
3   80603   89865   99033   157076  176543  177049  125111  181774
4   92842   92350   96164   165158  168580  191287  168710  192176
5   81002   90521   92902   165955  164884  185686  157606  188543
}\graphBtp

\pgfplotstableread{
A2ML    A2M     MinBFTL   MinBFT  MinZZL  MinZZ   CTL     CT      FBFTL   FBFT    FZZL    FZZ     PBFTL   PBFT    ZZL     ZZ        FPBFTOL FPBFTO  FZZOL   FZZO   
0.085   42166   0.074     47903   0.063   55303   0.083   43939   0.067   52646   0.056   60481   0.075   47552   0.057   60938     0.090   43744   0.070   46138
0.157   50060   0.106     72467   0.089   83867   0.148   53727   0.082   88649   0.068   106946  0.091   81161   0.065   110793    0.130   56314   0.140   57240
0.421   55502   0.292     80232   0.271   84612   0.406   58367   0.196   118784  0.158   143177  0.232   100346  0.156   144167    0.400   56174   0.410   59286
0.776   58735   0.556     85707   0.52    91003   0.732   61144   0.385   123084  0.289   160223  0.472   101282  0.291   161027    0.760   59671   0.790   60919
1.129   56451   0.77      82710   0.726   87159   1.078   59192   0.504   124711  0.399   158381  0.503   110946  0.367   171749    1.110   59454   1.120   61187
1.414   56753   0.971     82642   0.906   87404   1.338   59084   0.655   122392  0.467   157245  0.797   101131  0.467   170360    1.290   59875   1.260   61973
}\graphc

\pgfplotstableread{
X   A2M     MinBFT  	MinZZ   CT      FBFT    FZZ     PBFT    ZZ
1   1.078   0.887	0.833   1.013   0.85    0.820   0.942   0.661
2   1.045   0.912	0.875   1.034   0.891   0.853   0.95    0.694
3   1.082   0.941	0.921   1.05    0.919   0.885   0.982   0.736
4   1.116   0.928	0.921   1.077   0.923   0.884   1.011   0.772
5   1.111   0.921	0.92    1.06    0.928   0.874   0.989   0.759
6   1.12    0.938	0.93    1.07    0.925   0.879   0.948   0.742
}\graphDlt

\pgfplotstableread{
X   A2M     MinBFT  MinZZ   CT      FBFT    FZZ     PBFT    ZZ
1   51622   63011   65982   53953   64470   67214   60534   80590
2   52041   60372   62382   52768   62083   64082   59601   78841
3   50373   58717   59478   52689   60129   62357   57749   74363
4   49048   59058   59352   51377   59660   61841   56638   71536
5   50113   58354   59052   51953   59502   62590   57050   72044
6   49967   58234   58552   51142   59663   61964   57374   73181
}\graphDtp

\pgfplotstableread{
 X      MinZZ     FZZ     ZZ
 4      107596    134261    144932
 8      90140     100705    115504    
 16     66370     77673     86733
 24     47702     65556     66862
 32     39838     55382     55648       
}\graphEtp

\pgfplotstableread{
 X      MinZZ      FZZ       ZZ
 4      0.65       0.34      0.31
 8      0.8        0.38      0.34      
 16     1.09       0.63      0.47
 24     1.44       0.78      0.75
 32     1.78       0.92      0.85
}\graphElt

\newcommand{\graphATP}{
    \begin{tikzpicture}[plot]
        \begin{axis}[
                title={(\RN{2}) Scalability (Throughput)},
                scaled y ticks = false,
                xlabel={\axisFail},
                ylabel={\axisTP},
                ymin=0,
                ytick={0,40000,80000,120000,160000,200000},
                yticklabels={0,40K,80K,120K,160K,200K},
                xtick={4,8,16,24,32},
                xmode=log,
                log ticks with fixed point,
                legend to name={mainlegend},
                legend columns=-1,
            ]

            \addplot [black,every mark/.append style={solid,fill=\pgfplotsmarklistfill},mark=pentagon*] table[x={X},y={A2M}] {\graphAtp};
            \addplot [green,every mark/.append style={solid,fill=\pgfplotsmarklistfill},mark=square*] table[x={X},y={MinBFT}] {\graphAtp};
            \addplot [brown,every mark/.append style={solid,fill=\pgfplotsmarklistfill},mark=diamond*] table[x={X},y={MinZZ}] {\graphAtp};
            \addplot [blue!90!white!90,every mark/.append style={solid,fill=\pgfplotsmarklistfill},mark=o] table[x={X},y={CT}] {\graphAtp};
            \addplot [red,every mark/.append style={solid,fill=\pgfplotsmarklistfill},mark=halfdiamond*] table[x={X},y={FBFT}] {\graphAtp};
            \addplot [blue,every mark/.append style={solid,fill=\pgfplotsmarklistfill},mark=*] table[x={X},y={FZZ}] {\graphAtp};
            \addplot [teal,every mark/.append style={solid,fill=\pgfplotsmarklistfill},mark=triangle*] table[x={X},y={PBFT}] {\graphAtp};
            \addlegendimage {green!50!black!70,every mark/.append style={solid,fill=\pgfplotsmarklistfill},mark=halfcircle*}
            \addlegendimage {red,every mark/.append style={solid,fill=\pgfplotsmarklistfill},mark=halfcircle*}
            \addlegendimage {blue ,every mark/.append style={solid,fill=\pgfplotsmarklistfill,rotate=180},mark=diamond*}

            \addlegendentry [black,every mark/.append style={solid,fill=\pgfplotsmarklistfill},mark=pentagon*] \pbftea
            \addlegendentry [green,every mark/.append style={solid,fill=\pgfplotsmarklistfill},mark=square*] \MinBFT
            \addlegendentry [brown,every mark/.append style={solid,fill=\pgfplotsmarklistfill},mark=diamond*] \MinZZ
            \addlegendentry [blue!90!white!90,every mark/.append style={solid,fill=\pgfplotsmarklistfill},mark=o] \trustBFTc
            \addlegendentry [red,every mark/.append style={solid,fill=\pgfplotsmarklistfill},mark=halfdiamond*] \minLFT
            \addlegendentry [blue,every mark/.append style={solid,fill=\pgfplotsmarklistfill},mark=*] \minLFTZ
            \addlegendentry [teal,every mark/.append style={solid,fill=\pgfplotsmarklistfill},mark=triangle*] \pbft
            \addlegendentry [green!50!black!70,every mark/.append style={solid,fill=\pgfplotsmarklistfill},mark=halfcircle*] \ZZ
            \addlegendentry [red,every mark/.append style={solid,fill=\pgfplotsmarklistfill},mark=halfcircle*] {o\minLFT{}}
            \addlegendentry [blue ,every mark/.append style={solid,fill=\pgfplotsmarklistfill,rotate=180},mark=diamond*] {o\minLFTZ}

        \end{axis}
    \end{tikzpicture}
}

\newcommand{\graphALat}{
    \begin{tikzpicture}[plot]
        \begin{axis}[
                title={(\RN{3}) Scalability (Latency)},
                xlabel={\axisFail},
                ymin=0,
                ylabel={\axisLat},
                xtick={4,8,16,24,32},
                xmode=log,
                log ticks with fixed point,
            ]
            \addplot [black,every mark/.append style={solid,fill=\pgfplotsmarklistfill},mark=pentagon*] table[x={X},y={A2M}] {\graphAlt};
            \addplot [green,every mark/.append style={solid,fill=\pgfplotsmarklistfill},mark=square*] table[x={X},y={MinBFT}] {\graphAlt};
            \addplot [brown,every mark/.append style={solid,fill=\pgfplotsmarklistfill},mark=diamond*] table[x={X},y={MinZZ}] {\graphAlt};
            \addplot [blue!90!white!90,every mark/.append style={solid,fill=\pgfplotsmarklistfill},mark=o] table[x={X},y={CT}] {\graphAlt};
            \addplot [red,every mark/.append style={solid,fill=\pgfplotsmarklistfill},mark=halfdiamond*] table[x={X},y={FBFT}] {\graphAlt};
            \addplot [blue,every mark/.append style={solid,fill=\pgfplotsmarklistfill},mark=*] table[x={X},y={FZZ}] {\graphAlt};
            \addplot [teal,every mark/.append style={solid,fill=\pgfplotsmarklistfill},mark=triangle*] table[x={X},y={PBFT}] {\graphAlt};
        \end{axis}
    \end{tikzpicture}
}

\newcommand{\graphBTP}{
    \begin{tikzpicture}[plot]
        \begin{axis}[
                title={(\RN{4}) Batch Size (Throughput)},
                xlabel={\axisbatches},
                ymin=0,
                ylabel={\axisTP},
                ytick={0,40000,80000,120000,160000,200000},
                yticklabels={0,40K,80K,120K,160K,200K},
                xticklabels={10,100,500,1000,5000},
                xtick={1,2,3,4,5},
                log ticks with fixed point,
                scaled y ticks = false,
            ]
            \addplot [black,every mark/.append style={solid,fill=\pgfplotsmarklistfill},mark=pentagon*] table[x={X},y={A2M}] {\graphBtp};
            \addplot [green,every mark/.append style={solid,fill=\pgfplotsmarklistfill},mark=square*] table[x={X},y={MinBFT}] {\graphBtp};
            \addplot [brown,every mark/.append style={solid,fill=\pgfplotsmarklistfill},mark=diamond*] table[x={X},y={MinZZ}] {\graphBtp};
            \addplot [blue!90!white!90,every mark/.append style={solid,fill=\pgfplotsmarklistfill},mark=o] table[x={X},y={CT}] {\graphBtp};
            \addplot [red,every mark/.append style={solid,fill=\pgfplotsmarklistfill},mark=halfdiamond*] table[x={X},y={FBFT}] {\graphBtp};
            \addplot [blue,every mark/.append style={solid,fill=\pgfplotsmarklistfill},mark=*] table[x={X},y={FZZ}] {\graphBtp};
            \addplot [teal,every mark/.append style={solid,fill=\pgfplotsmarklistfill},mark=triangle*] table[x={X},y={PBFT}] {\graphBtp};
        \end{axis}
    \end{tikzpicture}
}

\newcommand{\graphBLat}{
    \begin{tikzpicture}[plot]
        \begin{axis}[
                title={(\RN{5}) Batch Size (Latency)},
                xlabel={\axisbatches},
                ymin=0,
                ylabel={\axisLat},
                xticklabels={10,100,500,1000,5000},
                xtick={1,2,3,4,5},
                log ticks with fixed point,
            ]
            \addplot [black,every mark/.append style={solid,fill=\pgfplotsmarklistfill},mark=pentagon*] table[x={X},y={A2M}] {\graphBlt};
            \addplot [green,every mark/.append style={solid,fill=\pgfplotsmarklistfill},mark=square*] table[x={X},y={MinBFT}] {\graphBlt};
            \addplot [brown,every mark/.append style={solid,fill=\pgfplotsmarklistfill},mark=diamond*] table[x={X},y={MinZZ}] {\graphBlt};
            \addplot [blue!90!white!90,every mark/.append style={solid,fill=\pgfplotsmarklistfill},mark=o] table[x={X},y={CT}] {\graphBlt};
            \addplot [red,every mark/.append style={solid,fill=\pgfplotsmarklistfill},mark=halfdiamond*] table[x={X},y={FBFT}] {\graphBlt};
            \addplot [blue,every mark/.append style={solid,fill=\pgfplotsmarklistfill},mark=*] table[x={X},y={FZZ}] {\graphBlt};
            \addplot [teal,every mark/.append style={solid,fill=\pgfplotsmarklistfill},mark=triangle*] table[x={X},y={PBFT}] {\graphBlt};
        \end{axis}
    \end{tikzpicture}
}

\newcommand{\graphDTP}{
    \begin{tikzpicture}[plot]
        \begin{axis}[
                title={(\RN{6}) Regions (Throughput)},
                xlabel={\axisRegions},
                ymin=0,
                ylabel={\axisTP},
                ytick={0,20000,40000,60000,80000,100000},
                yticklabels={0,20K,40K,60K,80K,100K},
                xticklabels={1,2,3,4,5,6},
                xtick={1,2,3,4,5,6},
                log ticks with fixed point,
                scaled y ticks = false,
            ]
            \addplot [black,every mark/.append style={solid,fill=\pgfplotsmarklistfill},mark=pentagon*] table[x={X},y={A2M}] {\graphDtp};
            \addplot [green,every mark/.append style={solid,fill=\pgfplotsmarklistfill},mark=square*] table[x={X},y={MinBFT}] {\graphDtp};
            \addplot [brown,every mark/.append style={solid,fill=\pgfplotsmarklistfill},mark=diamond*] table[x={X},y={MinZZ}] {\graphDtp};
            \addplot [blue!90!white!90,every mark/.append style={solid,fill=\pgfplotsmarklistfill},mark=o] table[x={X},y={CT}] {\graphDtp};
            \addplot [red,every mark/.append style={solid,fill=\pgfplotsmarklistfill},mark=halfdiamond*] table[x={X},y={FBFT}] {\graphDtp};
            \addplot [blue,every mark/.append style={solid,fill=\pgfplotsmarklistfill},mark=*] table[x={X},y={FZZ}] {\graphDtp};
            \addplot [teal,every mark/.append style={solid,fill=\pgfplotsmarklistfill},mark=triangle*] table[x={X},y={PBFT}] {\graphDtp};
        \end{axis}
    \end{tikzpicture}
}

\newcommand{\graphDLat}{
    \begin{tikzpicture}[plot]
        \begin{axis}[
                title={(\RN{7}) Regions (Latency)},
                xlabel={\axisRegions},
                ymin=0,
                ylabel={\axisLat},
                xticklabels={1,2,3,4,5,6},
                ytick={0,0.2,.4,.6,.8,1,1.2},
                xtick={1,2,3,4,5,6},
                log ticks with fixed point,
            ]
            \addplot [black,every mark/.append style={solid,fill=\pgfplotsmarklistfill},mark=pentagon*] table[x={X},y={A2M}] {\graphDlt};
            \addplot [green,every mark/.append style={solid,fill=\pgfplotsmarklistfill},mark=square*] table[x={X},y={MinBFT}] {\graphDlt};
            \addplot [brown,every mark/.append style={solid,fill=\pgfplotsmarklistfill},mark=diamond*] table[x={X},y={MinZZ}] {\graphDlt};
            \addplot [blue!90!white!90,every mark/.append style={solid,fill=\pgfplotsmarklistfill},mark=o] table[x={X},y={CT}] {\graphDlt};
            \addplot [red,every mark/.append style={solid,fill=\pgfplotsmarklistfill},mark=halfdiamond*] table[x={X},y={FBFT}] {\graphDlt};
            \addplot [blue,every mark/.append style={solid,fill=\pgfplotsmarklistfill},mark=*] table[x={X},y={FZZ}] {\graphDlt};
            \addplot [teal,every mark/.append style={solid,fill=\pgfplotsmarklistfill},mark=triangle*] table[x={X},y={PBFT}] {\graphDlt};
        \end{axis}
    \end{tikzpicture}
}

\newcommand{\graphC}{
    \begin{tikzpicture}[plot,scale=1.15]
        \begin{axis}[
                width=362.5pt,
                title={(\RN{1}) Throughput vs. Latency},
                scaled y ticks = false,
                xlabel={\axisLat},
                ylabel={\axisTP},
                ymin=0,
                ytick={0,40000,80000,120000,160000},
                yticklabels={0,40K,80K,120K,160K},
            ]
            \addplot [black,every mark/.append style={solid,fill=\pgfplotsmarklistfill},mark=pentagon*] table[x={A2ML},y={A2M}] {\graphc};
            \addplot [green,every mark/.append style={solid,fill=\pgfplotsmarklistfill},mark=square*] table[x={MinBFTL},y={MinBFT}] {\graphc};
            \addplot [brown,every mark/.append style={solid,fill=\pgfplotsmarklistfill},mark=diamond*] table[x={MinZZL},y={MinZZ}] {\graphc};
            \addplot [blue!90!white!90,every mark/.append style={solid,fill=\pgfplotsmarklistfill},mark=o] table[x={CTL},y={CT}] {\graphc};
            \addplot [red,every mark/.append style={solid,fill=\pgfplotsmarklistfill},mark=halfdiamond*] table[x={FBFTL},y={FBFT}] {\graphc};
            \addplot [blue,every mark/.append style={solid,fill=\pgfplotsmarklistfill},mark=*] table[x={FZZL},y={FZZ}] {\graphc};
            \addplot [teal,every mark/.append style={solid,fill=\pgfplotsmarklistfill},mark=triangle*] table[x={PBFTL},y={PBFT}] {\graphc};
            \addplot [red,every mark/.append style={solid,fill=\pgfplotsmarklistfill},mark=halfcircle*] table[x={FZZOL},y={FZZO}] {\graphc};
            \addplot [blue ,every mark/.append style={solid,fill=\pgfplotsmarklistfill,rotate=180},mark=diamond*] table[x={FPBFTOL},y={FPBFTO}] {\graphc};
        \end{axis}
    \end{tikzpicture}
}

\newcommand{\graphETP}{
    \begin{tikzpicture}[plot]
        \begin{axis}[
                scaled y ticks = false,
                xlabel={\axisFail},
                ylabel={\axisTP},
                ymin=0,
                ytick={0,40000,80000,120000,160000,200000},
                yticklabels={0,40K,80K,120K,160K,200K},
                xtick={4,8,16,24,32},
                xmode=log,
                log ticks with fixed point,
            ]

            \addplot [brown,every mark/.append style={solid,fill=\pgfplotsmarklistfill},mark=diamond*] table[x={X},y={MinZZ}] {\graphEtp};
            \addplot [red,every mark/.append style={solid,fill=\pgfplotsmarklistfill},mark=halfdiamond*] table[x={X},y={FBFT}] {\graphAtp};
            \addplot [blue,every mark/.append style={solid,fill=\pgfplotsmarklistfill},mark=*] table[x={X},y={FZZ}] {\graphAtp};
            \addplot [teal,every mark/.append style={solid,fill=\pgfplotsmarklistfill},mark=triangle*] table[x={X},y={PBFT}] {\graphAtp};
            \addplot [green!50!black!70,every mark/.append style={solid,fill=\pgfplotsmarklistfill},mark=halfcircle*] table[x={X},y={ZZ}] {\graphEtp};

        \end{axis}
    \end{tikzpicture}
}

\newcommand{\graphELat}{
    \begin{tikzpicture}[plot]
        \begin{axis}[
                xlabel={\axisFail},
                ymin=0,
                ylabel={\axisLat},
                xtick={4,8,16,24,32},
                xmode=log,
                ytick={0,.5,1,1.5,2,2.5},
                ymax=2,
                log ticks with fixed point,
            ]
            \addplot [brown,every mark/.append style={solid,fill=\pgfplotsmarklistfill},mark=diamond*] table[x={X},y={MinZZ}] {\graphElt};
            \addplot [red,every mark/.append style={solid,fill=\pgfplotsmarklistfill},mark=halfdiamond*] table[x={X},y={FBFT}] {\graphAlt};
            \addplot [blue,every mark/.append style={solid,fill=\pgfplotsmarklistfill},mark=*] table[x={X},y={FZZ}] {\graphAlt};
            \addplot [teal,every mark/.append style={solid,fill=\pgfplotsmarklistfill},mark=triangle*] table[x={X},y={PBFT}] {\graphAlt};
            \addplot [green!50!black!70,every mark/.append style={solid,fill=\pgfplotsmarklistfill},mark=halfcircle*] table[x={X},y={ZZ}] {\graphElt};
        \end{axis}
    \end{tikzpicture}
}

\newcommand{\graphPBFT}{
    \definecolor{bar-0}{RGB}{105,200,76}
    \definecolor{bar-1}{RGB}{215, 200, 76}
    \definecolor{bar-2}{RGB}{172, 179, 52}
    \definecolor{bar-3}{RGB}{250, 183, 51}
    \definecolor{bar-4}{RGB}{255, 142, 21}
    \definecolor{bar-5}{RGB}{255, 78, 17}
    \definecolor{bar-6}{RGB}{255, 13, 13}
    \begin{tikzpicture}[plot,scale=1]
        \begin{axis}[
                ybar,
                scaled y ticks = false,
                bar shift=0,
                ymin=0,
                bar width=15pt,
                ylabel={\axisTP},
                ytick={0,20000,40000,60000,80000,100000},
                yticklabels={0,20K,40K,60K,80K,100K},
                xticklabels={0,5,10,15,20,25,30},
                xticklabels={0,a,b,c,d,e,f,g},
            ]
            \addplot[style={fill=bar-0},error bars/.cd, y dir=both, y explicit,error bar style=black]
                coordinates   {(0,84212) += (0,1784) -= (0,1639)};
            \addplot[style={fill=bar-1},error bars/.cd, y dir=both, y explicit,error bar style=black] 
                coordinates   {(5,68799) += (0,1506) -= (0,2610)};
            \addplot[style={fill=bar-4},error bars/.cd, y dir=both, y explicit,error bar style=black] 
                coordinates   {(10,59302) += (0,3009) -= (0,1620)};
            \addplot[style={fill=bar-6},error bars/.cd, y dir=both, y explicit,error bar style=black] 
                coordinates   {(15,42704) += (0,1625) -= (0,1580)};
            \addplot[style={fill=bar-1},error bars/.cd, y dir=both, y explicit,error bar style=black] 
                coordinates   {(20,68059) += (0,1734) -= (0,2305)};
            \addplot[style={fill=bar-4},error bars/.cd, y dir=both, y explicit,error bar style=black] 
                coordinates   {(25,59391) += (0,1363) -= (0,1420)};
            \addplot[style={fill=bar-6},error bars/.cd, y dir=both, y explicit,error bar style=black] 
                coordinates   {(30,42127) += (0,1519) -= (0,2424)};
        \end{axis}
    \end{tikzpicture}
}

%% file: abstract.tex
The growing interest in reliable multi-party applications has 
fostered widespread adoption of Byzantine Fault-Tolerant (\BFT{}) consensus protocols.
%that can handle malicious attacks from byzantine replicas. 
Existing \BFT{} protocols need $\f$ more replicas than Paxos-style protocols 
to prevent equivocation attacks.
\trustBFT{} protocols instead seek to minimize this cost by making use of trusted components at replicas. 

This paper makes two contributions. 
First, we analyze the design of existing \trustBFT{} protocols and 
uncover three fundamental limitations that preclude most practical deployments. 
Some of these limitations are fundamental, while others  are linked to the state of trusted components today. 
Second, we introduce a novel suite of consensus protocols, \lftBFT{}, that 
attempts to sidestep these issues. 
We show that our \lftBFT{} protocols achieve up to $185\%$ more throughput than their \trustBFT{} counterparts.   

%% file: intro.tex
\section{Introduction}
\label{s:intro}
Byzantine Fault-Tolerant (\BFT{}) protocols allow multiple parties to perform shared computations and 
reliably store data without fully trusting each other; 
the system will remain correct even if a subset of participants behave maliciously~\cite{blockchain-book,pbftj,zyzzyva,hotstuff,geobft}. 
These protocols aim to ensure that all the 
replicas reach a {\em consensus} on the order of incoming client requests.
The popularity of these \BFT{} protocols is evident from the fact 
that they are widely used today, with applications ranging from safeguarding replicated and sharded databases~\cite{resilientdb,basil}, 
edge applications~\cite{wedgechain,resilient-edge}, and blockchain applications~\cite{bitcoin,ether,hyperledger-fabric}.

\BFT{} protocols tolerate a subset of participants behaving arbitrarily: 
a malicious actor can delay, reorder or drop messages ({\em omission faults}); 
it can also send conflicting information to participants ({\em equivocation}). 
As a result, \BFT{} consensus protocols are costly: 
maintaining correctness in the presence of such attacks requires a minimum of $3\f+1$ 
participants, where $\f$ participants can be malicious. 
This is in contrast to crash-fault tolerant protocols (\CFT{}), where participants can fail by crashing, 
which require $2\f+1$ participants only for correctness~\cite{paxos,raft}. 

To minimise this additional cost, some existing \BFT{} protocols leverage trusted components 
to curb the degree to which participants can behave arbitrarily~\cite{sgx,sanctum,keystone,occlum}. 
A trusted component provably performs a specific computation:
incrementing a counter~\cite{trinc,minbft}, appending to a log~\cite{a2m}, 
or more advanced options like executing a complex algorithm~\cite{arm-trustzone,iceclave,eactors}. 
While there exists a large number of \BFT{} protocols that leverage these 
trusted components~\cite{a2m,trinc,hotstuffm} 
(we refer to these protocol as \trustBFT{} protocols for simplicity), 
they all proceed in a similar fashion: 
they {\em force} each replica  
to commit to a single order for each request by 
having the trusted component sign each message it sends.
In turn, each trusted component either: 
(1) records the chosen order for a client request in an append-only log, or 
(2) binds this order for the request with the value of a monotonically increasing counter.
Committing to an order in this way allows these protocols to remain safe with 
$2\f+1$ replicas only, bringing them in line with their \CFT{} counterparts.

While reducing replication cost is a significant benefit, this paper argues that current 
\trustBFT{} protocols place {\em too much trust in trusted components}. 
Our analysis uncovers three fundamental issues with existing \trustBFT{} implementations that preclude most practical deployments:
(i) limited \revise{responsiveness} for clients, 
(ii) safety concerns associated with trusted components, and 
(iii) inability to perform multiple consensuses in parallel.

{\em \changebars{Responsiveness}{Loss of responsiveness with a single message delay.}}
We observe that \changebars{}{a single message delay is enough for} byzantine replicas
\changebars{can successfully}{to} prevent a client from receiving a response for its transactions. 
While the transaction 
\changebars{will still commit}{may appear committed} (\changebars{consensus}{replica} liveness), the system will appear to clients as stalled 
and thus appear non-responsive to clients.

\revise{
\trustBFT{} protocols allow a reduced quorum size of $\f+1$ to commit a request. 
As $\f$ of those may be Byzantine, only one honest replica is guaranteed to execute the operation. 
This is insufficient to guarantee that a client will receive the necessary $\f+1$ matching responses 
post operation execution and thus validate that the response is indeed valid.
}

{\em Loss of Safety under Rollback.}
Existing \trustBFT{} protocols consider an idealised model of trusted computations.
They assume that the trusted components cannot be compromised and 
that their data remains persistent in the presence of a malicious host.
\revise{
This assumption does not yet align with current hardware functionality. 
}
A large number of these protocols employ Intel SGX enclaves 
for trusted computing~\cite{damysus,hybster,ia-ccf}.
Unfortunately, SGX-based designs have been shown to suffer from rollback attacks~\cite{themis,rote,mem-attacks}, 
and the solutions to mitigate these attacks lack practical deployments~\cite{sgx-survey}.
\revise{
Hardware enclaves that do provably defend against rollback attacks, such as
}
persistent counters~\cite{sgx} and TPMs~\cite{tpm},
have prohibitively high latencies (tens of milliseconds)~\cite{adamcs,trinc,memoir}.

{\em Sequential Consensus.}
Existing \trustBFT{} protocols are inherently sequential as they
require each outgoing message to be ordered and attested by trusted components. While
recent work mitigates this issue by pipelining consensus phases~\cite{damysus} or
running multiple independent consensus invocations~\cite{hybster}, 
their performance remains fundamentally limited
\revise{
by the RTT of each protocol phase.
}
In fact, despite their lower replication factor, we observe that \trustBFT{} protocols achieve
lower throughput than traditional \changebars{parallel}{out-of-order} \BFT{} protocols~\cite{pbftj,geobft,basil,kauri}.

This paper argues that \trustBFT{} protocols have targeted the wrong metric: 
\revise{
while reducing the replication factor to $2\f+1$ may seem appealing from a resource efficiency or management overhead standpoint, 
it, paradoxically, comes at a significant performance cost. 
}
\changebars{Nonetheless, trusted components can}{Trusted components, however, can} still bring huge benefits to \BFT{} consensus \textit{when they use $3\f+1$ replicas}. 
We propose a novel suite of consensus algorithms ({\bf Flexi}ble {\bf Trust}ed \BFT{} (\lftBFT{})), 
which address the aforementioned limitations.  
These protocols are always \revise{responsive} and achieve high throughput as they 
(1) make minimal use of trusted components (once per \revise{client operation} and at the primary replica only), and 
(2) support parallel consensus invocations. 
Both these properties are made possible by the 
\revise{
ability to use large quorums (of size $2\f+1$) when using $3\f+1$ replicas.
}
Our techniques can be used to convert any \trustBFT{} protocol into a \lftBFT{} protocol. 
We provide as examples two such conversions: 
\minLFT{} and \minLFTZ{}, two protocols based on \pbft{}~\cite{pbftj}/ \MinBFT{}~\cite{minbft} and \ZZ{}~\cite{zyzzyva}/ \MinZZ~\cite{minbft}, respectively. 
\minLFT{} follows a similar structure to \pbft{}, but requires one less communication phase. 
\minLFTZ{} is, we believe, of independent interest: the protocol achieves consensus in a single linear phase 
without using expensive cryptographic constructs such as threshold signatures.
Crucially, unlike the \ZZ{} and \MinZZ{} protocols, \minLFTZ{} continues to commit in a single-round even when a 
single participant misbehaves, thus maintaining high-throughput~\cite{geobft,upright,aardvark}.
Further, \minLFTZ{} does not face the safety bug like \ZZ{}~\cite{zyzzyva-unsafe}.

We evaluate our \lftBFT{} variants against {\em five} protocols: 
{\em three} \trustBFT{} and {\em two} \BFT{} systems. 
For fair evaluation, we implement these protocols on the open-source 
\ResilientDB{} fabric~\cite{geobft,rcc,kauri}.
Our evaluation on a real-world setup of $97$ replicas and up to \SI{80}{k} clients
shows that our \lftBFT{} protocols achieve up to $185\%$ and $100\%$ more
throughput than their \trustBFT{} and \BFT{} counterparts, respectively.
Further, we show that our \lftBFT{} protocols continue outperforming 
these protocols even when replicas are distributed across {\em six} locations in four continents.
In summary, we make the following {\em four} contributions:
\begin{itemize}[nosep]
\item \revise{We identify a responsiveness issue in existing \trustBFT{} protocols when an honest replica faces 
temporary message delays.}

\item 
We highlight that rollback attacks on trusted components limit practical deployments.

\item 
We observe that existing \trustBFT{} protocols are inherently sequential. The lack of support for parallel
        consensus invocations artificially limits throughput compared to traditional \BFT{} prtoocols.

\item We present \lftBFT{} protocols: a novel suite of protocols that 
support concurrent consensus invocations and require minimal access to the trusted components.

\end{itemize}

%% file: prelim.tex
\section{System model and notations}\label{s:model}
We adopt the standard communication and failure model adopted by most \BFT{} protocols~\cite{poe,pbftj,geobft,basil,a2m,kauri}, including
all existing \trustBFT{} protocols~\cite{a2m,minbft,hotstuffm,damysus,hybster}.

We consider a replicated service $\Service$ consisting of 
set of $n$ replicas $\Replicas$ of which at most $f$ can behave arbitrarily. 
The remaining $n-f$ are honest: they will follow the protocol and remain live. 
We \changebars{additionally assume}{further consider} the existence of a finite set of clients $\Clients$ of which arbitrarily many
can be malicious.

Similarly, we inherit standard communication assumptions: we assume the
existence of \textit{authenticated channels} (Byzantine replicas can impersonate
each other but no replica can impersonate an honest replica) and 
standard cryptographic primitives such as MACs and digital signatures (\DS{}). 
We denote a message $m$ signed by a replica $\Replica{}$ using \Name{DS} as
{\bf $\SignMessage{m}{\Replica{}}$}.
We employ a \emph{collision-resistant} hash function
$\Hash{\cdot}$ to map an arbitrary value $v$ to a constant-sized digest $\Hash{v}$.
We adopt the same partial synchrony model adopted
in most \BFT{} systems: safety is guaranteed in an asynchronous environment 
where messages can get lost, delayed, or duplicated.  
Liveness, however, is only guaranteed during periods of synchrony~\cite{pbftj,geobft,basil,a2m,hotstuff}.
Each replica only accepts a message if it is {\bf \em well-formed}. 
A well-formed message has a valid signature and passes all the protocol checks.  

\revise{
We follow Schneider's seminal work~\cite{fred-smr} to distinguish between consensus and Replicated State Machine (\RSM).
A consensus protocol aims to order the operations among the participants, 
whereas the correctness of an \RSM{} is not just defined by the agreement but also 
that the response to a client's transaction $\Transaction{}$ is the result of applying each transaction that precedes $\Transaction{}$ in order.
This description allows us to define the following guarantees:

\begin{description}
\item[\bf Consensus Safety.] 
If two honest replicas $\Replica{1}$ and $\Replica{2}$ order a transaction $\Transaction{}$ at sequence numbers $k$ and 
$k'$, then $k = k'$.

\item[\bf Consensus Liveness.]
If a honest replica commits $\Transaction{}$, then all honest replicas eventually commit $\Transaction{}$.

\item[\bf \RSM{} Safety.]
Given a consensus order $O$ of transactions and a transacton $\Transaction{}$, 
the output of $\Transaction{}$ is consistent with applying $\Transaction$ after applying all the transactions in $O$ 
according to the semantic characterisation of the state machine.

\item[\bf \RSM{} Liveness.]
If a client sends a transaction $\Transaction{}$, then it will eventually receive a response for $\Transaction{}$.
\end{description}
}

We additionally assume that our replicated service $\Service$ includes
a set of trusted components. 
These trusted components offer the following abstraction:

\begin{definition} \label{def:trust-comp}
A trusted component $\Trust{}$ is a cryptographically 
secure entity, which has a negligible probability of being 
compromised by byzantine adversaries. 
$\Trust{}$ provides access to a function \foo{} and, 
when called, always computes \foo{}.
\end{definition}

Existing \trustBFT{} protocols 
assume that each replica $\Replica \in \Replicas$ 
has access to a {\em co-located} trusted component. 
We use the notation $\Trust{\Replica}$ to denote the trusted component at 
the ``host'' replica $\Replica$.
As $\Trust{\Replica}$ computes \foo{} and $\Trust{\Replica}$ cannot be 
compromised by the host $\Replica$, existing \trustBFT{} protocols claim integrity 
in the computation of \foo{}.

%% file: back.tex
\section{Primer on \BFT{} Consensus}
\label{s:back}

To highlight the limitations of \trustBFT{} protocols, we first explain \BFT{} consensus. 
To this effect, we summarize the structure of \pbft{}-like systems,
which represent most modern \BFT{} protocols today~\cite{zyzzyva,sbft,poe,hotstuff,geobft,kauri}. 
For simplicity of exposition, we focus specifically on \pbft{}~\cite{pbftj}.

\pbft{} adopts the system and communication model previously described and guarantees safety and liveness for $\n=3\f+1$ replicas
where at most $\f$ can be Byzantine. The protocol follows a primary-backup model: one replica is designated as the primary (or leader), 
while the remaining replicas act as backups. 
At a high-level, all \BFT{} protocols consist of two logical phases: an agreement phase where replicas agree to commit a specific operation,
and a second durability phase where this decision will persist even when the leader is malicious and replaced as part of a \textit{view-change} (view-changes
are the process through which leaders are replaced).  
For ease of understanding, we focus on explaining the simplest failure-free run of \pbft{}.

\revise{
{\bf Client Library.} 
The input to any \RSM{} is a client request. 
Generally, a client $\Client$ invokes the {\em client side library} when it 
wants the \RSM{} to process its transaction $\Transaction{}$.
To do so, $\Client{}$ issues a signed message  
$\SignMessage{\Transaction}{\Client{}}$ to the library, which forwards the message 
to the \RSM's primary replica $\Primary{}$.
The client library at $\Client{}$ waits for identical responses from 
$\f+1$ replicas before returning the result \Result{} to $\Client$.
Waiting for $\f+1$ matching responses
ensures execution correctness as at least one honest replica can vouch for that result.
}

\begin{enumerate}[wide,nosep]
\item {\bf Pre-prepare.} 
When the primary replica $\Primary{}$ receives a well-formed client request $m$, it assigns the corresponding transaction
a new sequence number $k$ and sends a $\Name{Preprepare}$ message to all backups.

\item {\bf Prepare.}
When a replica $\Replica{} \in \Replicas{}$  receives, for the first time, a  
well-formed $\Name{Preprepare}$ message from $\Primary{}$ for a given sequence number $k$, $\Replica{}$ broadcasts to all 
other replicas a $\Name{Prepare}$ message agreeing to support this (sequence number, transaction) pairing.

\item {\bf Commit.}
When a replica $\Replica{}$ receives identical $\Name{Prepare}$ messages from $2\f+1$ replicas, 
it marks the request $m$ as {\em prepared} and broadcasts a $\Name{Commit}$ message to all replicas. 
When a request is prepared, a replica has the guarantee that no conflicting 
request $m'$ for the same sequence number will ever be prepared by this leader.

\item {\bf Execute.}
Upon receiving $2\f+1$ matching $\Name{Commit}$ messages, $\Replica$ marks the request as {\em committed} and
executable. 
Each replica $\Replica$ executes every request in sequence number order: the node waits until request for slot $k-1$ has successfully
executed before executing the transaction at slot $k$. 
Finally, $\Replica$ returns the result $\Result$ of the operation to the client $\Client$. 
\end{enumerate}

%% file: trusted.tex
\section{Trusted BFT Consensus}
\label{s:trust}

\BFT{} protocols successfully implement consensus, but at a cost: they require a higher
replication factor ($3\f+1$) and, in turn, must process significantly more messages 
(all of which must be authenticated and whose signatures/MACs must be verified). 
Equivocation is the primary culprit: 
a byzantine replica can tell one group 
of honest replicas that it plans to order a transaction $\Transaction{}$ 
before transaction $\Transaction{'}$, and tells another group of 
replicas that it will order $\Transaction{'}$ before $\Transaction{}$.

In a \CFT{} system, it is sufficient for
quorums to intersect in a single replica to achieve \textit{agreement}. 
In \BFT{} systems, as byzantine replicas can \textit{equivocate}, quorums must instead intersect in
one {\em honest} replica (or, in other words, must intersect in at least $\f+1$ replicas). 

To address this increased cost, \trustBFT{} protocols make use of trusted components 
(such as Intel SGX, AWS Nitro) at each node~\cite{equivocation-trusted, a2m,attestation-mechanisms}. 
Trusted components {\em cannot}, by assumption, be compromised by malicious actors. 
They can thus be used to prevent replicas from equivocating. 
{\em In theory}, this should be sufficient to safely convert any \BFT{} protocol into a \CFT{} system~\cite{equivocation-trusted}. 
In practice, however, as we highlight in this work, this process is less than straightforward. 

\subsection{Trusted Component Implementations}
\revise{
The easiest option is to run the full \BFT{} consensus inside the trusted component~\cite{spons-shields,rktio,sgxlkl,scone,eleos,palaemon}. 
However, this approach violates the principle of least privilege~\cite{least-privilege,glamdring}.
}

Instead, existing protocols choose to put the smallest amount of computation inside the trusted component, 
which has the benefit of allowing custom hardware implementations that are more secure~\cite{a2m,trinc}. 
There are two primary approaches: append-only logs, and monotonically increasing counters. 
Careful use of these primitives allows prior work to reduce the replication factor from $3\f+1$ to $2\f+1$.

{\bf Trusted Logs.}
\trustBFT{} protocols like \pbftea{}~\cite{a2m} and \hotstuffm~\cite{hotstuffm}
maintain, in each trusted component $\Trust{\Replica}$, a set of append-only logs.
Each log $\tid$ has a set of slots. 
We refer to each slot using an identifier $\tct$. 
Each trusted component $\Trust{\Replica}$ offers the following API. 

\begin{enumerate}[nosep]
\item {\bf $\Message{Append}{\tid,\knew,x}$} --
    Assume the last append to log $\tid$ of trusted component $\Trust{\Replica}$ was at slot $\tct$, then
	\begin{itemize}[nosep,wide]
	\item If $\knew =~\perp$, no slot location is specified, 
        $\Trust{\Replica}$ sets $\tct$ to $\tct+1$ and appends the value $x$ to slot $\tct$. 	
	\item If $\knew > \tct$, $\Trust{\Replica}$ appends $x$ to slot $\knew$ 
        and updates $\tct$ to $\knew$. 
	The slots in between can no longer be used.
	\end{itemize}	

\item {\bf $\Message{Lookup}{\tid,\tct}$} --  
If there exists a value $x$ at slot $\tct$ in log $\tid$ of $\Trust{\Replica}$, 
it returns an attestation 
{\bf $\SignMessage{\Message{Attest}{\tid,\tct,x}}{\Trust{\Replica}}$}.
\end{enumerate}
The $\MName{Append}$ function ensures that no two requests are ever logged 
at the same slot; 
the $\MName{Lookup}$ function confirms this fact
with a digitally-signed assertion $\SignMessage{\Message{Attest}{\tid,\tct,x}}{\Trust{\Replica}}$ 
where $\tct$ is the log position and $x$ is the stored message.

{\bf Trusted Counters.}
A separate line of work further restricts the scope of the trusted components. 
Rather than storing an attested log of messages, 
it simply stores a set of monotonically increasing counters~\cite{minbft,cheapbft}. 
These counters do not provide a $\MName{Lookup}$ function as they do not store messages. 
Instead, the $\Message{Append}{\tid,\knew,x}$ function, 
at the time of invocation, returns the attestation proof $\SignMessage{\Message{Attest}{\tid,\tct,x}}{\Trust{\Replica}}$, 
which states that the $\tid$-th counter updates its current value $\tct$ to $\knew$ and
bounds the updated value $\tct$ to message $x$.

These two designs are not mutually exclusive; protocols like
\Trinc{}~\cite{trinc}, \Hybster~\cite{hybster} and \Damsys~\cite{damysus}
require their trusted components to support both logs and counters,
where logs record the last few client requests.
\vspace{-4mm}

\subsection{\trustBFTb{} protocol}
\label{ss:tbft-protocol}
Existing \trustBFT{} protocols expect that in a system of $\n = 2\f+1$ replicas at most $\f$ replicas are byzantine.
Most \trustBFT{}~\cite{a2m,minbft,cheapbft,trinc,hybster,hotstuffm} follow a similar design, based on the
\pbftea~\cite{a2m} protocol; we describe it next. \pbftea{} is derived from \pbft{} and makes use
of a set of trusted logs.

{\bf \pbftea{} protocol steps.} 
In the \pbftea{} protocol, $\Trust{\Replica}$ stores  
{\em five} distinct sets of logs, corresponding to each phase of the consensus.
When the primary $\Primary{}$ receives a client request $m$, 
it calls the $\Message{Append}{\tid,k,m}$ function to assign $m$ a sequence number $k$.  
$\Trust{\Primary{}}$ logs $m$ in the $\tid$-th
{\em preprepare log} and returns an attestation 
$\SignMessage{\Message{Attest}{\tid,k,m}}{\Trust{\Primary{}}}$, 
which $\Primary$ forwards to all replicas along with the 
$\MName{Preprepare}$ message. 
The existence
of a trusted log precludes a primary from equivocating and sending
two conflicting messages $m$ and $m'$ for the same sequence number $k$.
When a replica $\Replica{}$ receives a well-formed $\MName{Preprepare}$ message for slot $k$, 
it creates a $\MName{Prepare}$ message $m'$, and 
calls the $\MName{Append}$ function on its $\Trust{\Replica{}}$ to log $m'$.
As a result, $\Trust{\Replica{}}$ logs $m'$ in its $\tid$-th {\em preprepare log} and 
returns an attestation $\SignMessage{\Message{Attest}{\tid,\tct,m'}}{\Trust{\Replica}}$, which 
$\Replica{}$ forwards along with the $\MName{Prepare}$ message to all replicas.

Once $\Replica{}$ receives $\f+1$ identical $\MName{Prepare}$ messages
from distinct replicas (including itself), it declares the transaction prepared.
Following this, $\Replica{}$ repeats the process by creating a $\MName{Commit}$ message (say $m''$)
and asking its $\Trust{\Replica{}}$ to log this message at slot $k$ in its
$\tid$-th \textit{prepare log}. 
$\Replica{}$ broadcasts the signed attestation 
$\SignMessage{\Message{Attest}{\tid,\tct,m''}}{\Trust{\Replica}}$ along with the $\MName{Commit}$ message.
Once $\Replica$ receives $\f+1$ matching $\MName{commit}$ messages for $m$ at slot $k$, it marks the operation committed.
\revise{
$\Replica{}$ executes $m$ once it has executed the request at slot $k-1$
and sends the result of execution (\Result). 
The client library returns \Result{} to the client when 
it receives $\f+1$ identical responses.
}

\revise{
{\bf \MinBFT{} and \MinZZ{} protocols.}
\MinBFT~\cite{minbft} improves over \pbftea{} by observing that the use of trusted components makes the 
$\MName{Commit}$ phase redundant. 
\MinBFT{} allows a replica to mark transactions as committed once it receives $\f+1$ identical $\MName{Prepare}$ messages.
\MinZZ~\cite{minbft} makes a similar observation to improve the \ZZ~\cite{zyzzyva} protocol: 
it allows replicas to speculatively execute an operation once they receive a $\MName{Preprepare}$ message from the primary.
However, unlike \ZZ{} where the client needs identical responses from $\n = 3\f+1$ replicas to mark its transaction as complete, 
in \MinZZ{} the client requires $\n = 2\f+1$ responses.
}

{\bf Checkpoints.}
Like \BFT{} protocols, \trustBFT{} protocols also periodically share checkpoints.
These checkpoints reflect the state of a replica $\Replica$'s trusted component $\Trust{\Replica}$ 
and enable log truncation.
During the checkpoint phase, $\Replica{}$ sends $\MName{Checkpoint}$ 
messages that include all the requests committed since the last checkpoint.
If $\Trust{\Replica}$ employs trusted logs, then it also provides attestations for each logged request. 
If instead $\Trust{\Replica}$ employs only trusted counters, $\Trust{\Replica}$ 
provides an attestation on the current value of its counters.
Each replica $\Replica{}$ marks its checkpoint as stable if it receives $\MName{Checkpoint}$
messages from $\f+1$ other replicas possibly including itself.

{\bf View Change.} 
The existence of trusted logs/counters precludes the primary from equivocating,
but it can still deny service and maliciously delay messages. The \textit{view-change} mechanism enables replicas
to replace a faulty primary; a view refers to the duration for which a specific replica was leader.
All \trustBFT{} protocols provide such a mechanism~\cite{a2m,cheapbft}: 
a view change is triggered when $\f+1$ replicas send a \PName{Viewchange} message. 
Requiring at least $\f+1$ messages
ensures that malicious replicas cannot alone attempt to replace leaders. Similar to the common-case protocol
$\f+1$ replicas must participate in every view-change quorum.

\begin{figure*}[t]
\centering
\footnotesize
\begin{tabular}{!{\vrule width 1.5pt}c|C{2.3cm}|C{1.2cm}|C{1.45cm}|C{2.0cm}|C{2.0cm}|C{4.6cm}!{\vrule width 1.5pt}} \HLine{1.5pt}
\rowcolor{black!15} Replicas & Trusted		& Liveness as \BFT{} 	& Out-of-Order	& Memory	& Only Primary needs active TC & Protocols \\ \HLine{1.5pt}

\multirow{3}{2.5em}{$2\f+1$} & Log		& \xmark	& \xmark	& High		& \xmark & \pbftea{}, \hotstuffm \\ \cline{2-7}

			 & Counter + Log	& \xmark	& \xmark	& Order of Log-size 	& \xmark & \Trinc, \Hybster, \Damsys{} \\ \cline{2-7}

			 & Counter		& \xmark	& \xmark	& Low		& \xmark & \MinBFT{}, \MinZZ{}, \CheapBFT{} \\ \HLine{1.5pt}

\rowcolor{green} $3\f+1$ & Counter		& \cmark	& \cmark	& Low		& \cmark & \lftBFT{} (\minLFT{}, \minLFTZ{}) \\ \HLine{1.5pt}
\end{tabular}
\caption{Comparing \trustBFT{} protocols. %across a variety of parameters.
From left to right: 
\Col{1}: type of trusted abstraction;
\Col{2}: identical liveness guarantees as \BFT{} protocols;
\Col{3}: support for out-of-order consensuses; 
\Col{4}: amount of memory needed; and
\Col{5}: only primary replica requires active trusted component.
\lftBFT{} are our proposals, which we present in this paper.
}
\label{fig:comp-protocols}
\end{figure*}

This design summarizes, to the best of our knowledge, most existing \trustBFT{} protocols, 
which adopt a similar structure. We summarize other approaches in Figure~\ref{fig:comp-protocols} 
and explain them in detail in Section~\ref{s:related}.

\begin{figure*}[t]
    \centering
    \begin{tikzpicture}[yscale=0.25,xscale=0.95]
        \draw[thick,draw=black!75] (0.75, 11.5) edge[green!50!black!90] ++(8.5, 0)
                                   (0.75,   4) edge ++(8.5, 0)
                                   (0.75,   5) edge ++(8.5, 0)
                                   (0.75,   6.5) edge ++(8.5, 0)
                                   (0.75,   7.5) edge ++(8.5, 0)
                                   (0.75,   9) edge ++(8.5, 0)
                                   (0.75,   10) edge[red] ++(8.5, 0);

        \draw[thin,draw=black!75] (1, 4) edge ++(0, 7.5)
                                  (2, 4) edge ++(0, 7.5)

				  (3.5, 4) edge ++(0, 7.5)
                                  (5, 4) edge ++(0, 7.5);
                                  %(6.5, 4) edge ++(0, 7.5);

	\draw[thick,dotted,draw=black!75] (5.7, 4) edge ++(0, 7.5);
	\draw[thick,dotted,draw=black!75] (6.5, 4) edge ++(0, 7.5);
	\draw[thin,draw=black!75] (7.5, 4) edge ++(0, 7.5);
	\draw[thin,draw=black!75] (8.5, 4) edge ++(0, 7.5);

	\node[left] at (0.8, 4) {\scriptsize $\Trust{\Replica{2}}$};
        \node[left] at (0.8, 5) {\scriptsize $\Replica{2}$};
	\node[left] at (0.3, 5) {\scriptsize $D = $};

	\node[left] at (0.8, 6.5) {\scriptsize $\Trust{\Replica{1}}$};
        \node[left] at (0.8, 7.5) {\scriptsize $\Replica{1}$};
	\node[left] at (0.3, 7.5) {\scriptsize $\Replica{} = $};

	\node[left] at (0.8, 9) {\scriptsize $\Trust{\Primary}$};
        \node[left,red] at (0.8, 10) {\scriptsize $\Primary$};
	\node[left,red] at (0.3, 10) {\scriptsize $\Faulty{} = $};
        \node[left] at (0.8, 11.5) {\scriptsize $\Client{}$};

	\draw[brown] (0.2,8.5) rectangle (0.8,10.4);
	\draw[brown] (0.2,6.0) rectangle (0.8,7.9);
	\draw[brown] (0.2,3.5) rectangle (0.8,5.4);

        \path[->] (1, 11.5) edge node[above,pos=0.8] {\small {$\Transaction$}} (2, 10)
		  (2, 10) edge (2.5, 9)
		  (2.5, 9) edge (3, 10)

                  (3, 10) edge (3.5, 7.5)
                          %edge (3.5, 5)

		  %Prepare                           
		  (3.5, 7.5) edge (4, 6.5)
		  (4.0, 6.5) edge (4.5, 7.5)

                  (4.5, 7.5) edge (5.0, 10)
			     edge (4.7, 5.8)

		  %Reply
                  (5.0, 7.5) edge (5.7, 11.5)

		  %Resend
		  (6.5, 11.5) edge node[above,pos=0.8] {\small {$\Transaction$}} (7.5, 10)
			      edge (7.5, 7.5)
			      edge (7.5, 5)

		  %Resend
		  (7.5, 7.5) edge node[green!50!black!90,above,pos=0.6,rotate=50] {\scriptsize \Name{Resend}}(8.5, 11.5)

		  (7.5, 5) edge (8.5, 10)
		  (7.5, 5) edge node[blue,align=left] {\scriptsize \Name{view} \\ \scriptsize \Name{change}}(8.5, 7.5)
                           ;

	% Delayed edges
	\draw[red,thick] (4.55,6.1) edge (4.75,5.8);
	\draw[red,thick] (4.55,5.8) edge (4.75,6.1);
	\node[red,below] at (4.35, 6.2) {\scriptsize \strut \Name{Delayed}};

        \node[below] at (2.8, 4.2) {\scriptsize \strut \Name{Preprepare}};
        \node[below] at (4.4, 4.2) {\scriptsize \strut\Name{Prepare}};
	%\node[below] at (5.8, 4.2) {\scriptsize \strut\Name{Commit}};
	\node[below] at (5.4, 4.2) {\scriptsize \strut\Name{Reply}};
	\node[below] at (6.2, 4.2) {\scriptsize \strut\Name{Wait}};
	\node[below] at (7.2, 4.2) {\scriptsize \strut\Name{Complain}};
	\node[above] at (4.5, 11) {\revise{{\scriptsize \strut \MinBFT{}}}};

	\node[below,red] at (8.9, 10.3) {\scriptsize \strut\Name{Ignore}};
	\node[below,brown,align=left] at (8.9, 7.8) {\scriptsize \strut $< \f+1$};

	%%%% PBFT
	\draw[thick,draw=black!75] (12.25, 11.5) edge[green!50!black!90] ++(6.15, 0)
                                   (12.25,   7) edge ++(6.15, 0)
                                   (12.25,   8) edge ++(6.15, 0)
                                   (12.25,   9) edge ++(6.15, 0)
                                   (12.25,   10) edge[red] ++(6.15, 0);

        \draw[thin,draw=black!75] (12.5, 7) edge ++(0, 4.5)
                                  (13, 7) edge ++(0, 4.5)
				  (14, 7) edge ++(0, 4.5)
                                  (15.5, 7) edge ++(0, 4.5)
                                  (17, 7) edge ++(0, 4.5);

        \node[left] at (12.4, 7) {\scriptsize $\Replica_3$};
	\node[left] at (12, 7) {\scriptsize $D = $};

        \node[left] at (12.4, 8) {\scriptsize $\Replica_2$};
        \node[left] at (12.4, 9) {\scriptsize $\Replica_1$};
	\node[left] at (12, 8.5) {\scriptsize $\Replica{} = $};
	\draw[decorate,decoration = {brace},thick] (12,7.7) --  (12,9.3);

        \node[left,red] at (12.3, 10) {\scriptsize $\Primary$};
	\node[left,red] at (12, 10) {\scriptsize $\Faulty{} = $};
        \node[left] at (12.3, 11.5) {\scriptsize $\Client{}$};

        \path[->] (12.5, 11.5)  edge node[above,pos=0.8] {\scriptsize $\Transaction$} (13, 10)
                  (13.0, 10)    edge (14, 9)
                                edge (14, 8)
                           
                  (14.0, 9)     edge (15.5, 10)
			        edge (15.5, 8)
			        edge (15.05, 7.5)

		  (14.75, 8)    edge (15.5, 10)
			        edge (15.5, 9)
			        edge (15.05, 7.5)

		  (15.5, 10)    edge (17, 9)
			        edge (17, 8)

		  (15.5, 9)     edge (17, 10)
			        edge (17, 8)
			        edge (16.55, 7.5)

		  (15.5, 8)     edge (17, 10)
			 	edge (17, 9)
			 	edge (16.55, 7.5)

                  (17, 9) edge (18, 11.5)
                  (17, 8) edge (18, 11.5)
                           ;

	% Delayed edges
	\draw[red,thick] (14.85,7.8) edge (15.15,7.4);
	\draw[red,thick] (14.85,7.4) edge (15.15,7.8);
	\node[red,below] at (14.45, 8.1) {\scriptsize \strut \Name{Delayed}};

	\draw[red,thick] (16.35,7.8) edge (16.65,7.4);
	\draw[red,thick] (16.35,7.4) edge (16.65,7.8);
	\node[red,below] at (15.95, 8.1) {\scriptsize \strut \Name{Delayed}};

        \node[below] at (13.5, 6.7) {\scriptsize \strut \Name{Preprepare}};
        \node[below] at (14.8, 6.7) {\scriptsize \strut\Name{Prepare}};
	\node[below] at (16.3, 6.7) {\scriptsize \strut\Name{Commit}};
	\node[below] at (17.5, 6.7) {\scriptsize \strut\Name{Reply}};
	\node[above] at (15, 11) {\scriptsize \strut \pbft{}};

    \end{tikzpicture}
    \caption{Disruption in service in \MinBFT{} protocol due to weak quorums in comparison to the \pbft{} protocol (Section~\ref{s:live}).
	}
    \label{fig:weak-quorum}
\end{figure*}
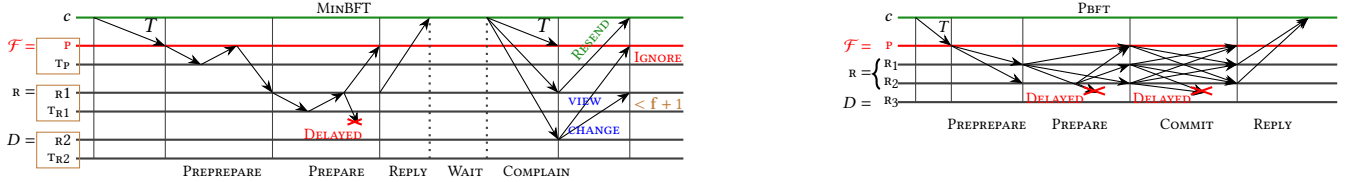

{\bf Challenges.}
\trustBFT{} protocols aim to provide the same safety and liveness guarantees as their 
\BFT{} counterparts.
However, we identify several {\em challenges} with their designs.
We observe that these protocols:  
(i) offer limited liveness for clients, 
(ii) encounter safety concerns associated with trusted components, and 
(iii) disallow consensus invocations from proceeding in parallel.
We will expand on these in upcoming sections.
To address these challenges, we present the design of our novel \lftBFT{} protocols in Section~\ref{s:extend}.

%% file: liveness.tex
\section{Restricted Responsiveness}
\label{s:live}
\changebars{We first observe that current \trustBFT{} protocols lack guaranteed client responsiveness}{
We first identify a liveness issue in current \trustBFT{} protocols}: 
\revise{
temporary delaying the messages of a single honest replica is
} 
sufficient to prevent a client from receiving a result for its transaction until the next checkpoint.
For any production-scale system, responsiveness (or {\em client liveness}) is as important a metric as {\em replica liveness}
(the operation committing at the replica). 
Specifically, \trustBFT{} protocols \revise{{\em as currently implemented}} guarantee that a quorum of $f+1$ replicas will commit a request 
\revise{(consensus liveness)}, but do not guarantee that
sufficiently many honest replicas will actually \textit{execute it}. 
These protocols cannot guarantee that the clients will receive $f+1$ identical
responses from distinct replicas \revise{(\RSM{} liveness)}, which is the necessary threshold to validate execution results.
The ease with which such a \revise{\RSM{} liveness (or client responsiveness)} issue can be triggered highlights the brittleness of current \trustBFT{} approaches.
We highlight this issue using \revise{\MinBFT{}} as an example, but, to the best of our knowledge, the same problem arises with all existing \trustBFT{} protocols.

\begin{claim}\label{cl:lack-liveness}
In a replicated system $\Service$ of replicas $\Replicas$ and
clients $\Clients$,
where $\abs{\Replicas} = \n = 2\f+1$, 
\revise{
there exist an execution in which client responsiveness is not guaranteed 
}
\end{claim}

\begin{proof}
Assume a run of the \revise{\MinBFT{}} protocol (Figure~\ref{fig:weak-quorum}).  
We know that $\f$ of the replicas in $\Replicas{}$ are byzantine. 
We represent them with set $\Faulty$.
The remaining $\f+1$ replicas are honest. 
Let us distribute them  into two groups:
$D$ and $\Replica{}$, such that
$\abs{D} = \f$ replicas and $\Replica{}$ be the remaining replica.

Assume that the primary $\Primary{}$ is byzantine ($\Primary \in \Faulty$) and 
all replicas in $\Faulty$ intentionally fail to send replicas in $D$ any messages.
As is possible in a partially synchronous system, we further assume that the \revise{$\PName{Prepare}$ messages} from the replica $\Replica$ 
to those in $D$ are delayed.

$\Primary{}$ sends a $\MName{Preprepare}$ message for a client $\Client$'s transaction $\Transaction$ 
to all replicas in $\Faulty$ and $\Replica$. 
These $\f+1$ replicas are able to \revise{prepare $\Transaction$}.
All messages from replica $\Replica$ to those in $D$ are delayed.
It follows that $\Replica{}$ is the only honest replica to receive the transaction and 
reply to the client $\Client$. 
Replicas in $\Faulty{}$ in contrast, fail to respond.
Unfortunately, the client needs $\f+1$ responses to validate the correctness of the executed operation, and thus cannot make progress.

Eventually, the client will complain to all the replicas that it has not received sufficient responses 
for its transaction $\Transaction{}$.
Having not heard from the leader about $\Transaction{}$, replicas in $D$ will eventually vote to trigger a view-change. 
Unfortunately, a view-change requires
at least $\f+1$ votes to proceed; otherwise, byzantine replicas could stall system progress by 
constantly triggering spurious view-changes.
The system can thus no longer successfully execute operations: 
the client will never receive enough matching responses, and no view-change
can be triggered to address this issue.
\end{proof}

This attack is not specific to \revise{\MinBFT{}}
and applies to other protocols like \revise{\pbftea{}}, \Trinc{}, \CheapBFT{}, and \Hybster{} 
as they have similar consensus phases and commit rules. It also applies
to streamlined protocols \hotstuffm{} and \Damsys{} which frequently rotate primaries.

{\bf Weak Quorums.}
\revise{
The smaller set of replicas in \trustBFT{} protocols triggers this responsiveness issue.
In \trustBFT{} protocols, a quorum of $\f+1$ matching votes suffices to enforce consensus safety as 
trusted components certify the position of the transaction in the log and preclude equivocation. 
Unfortunately, we observe that the quorums of $\f+1$ replicas are insufficient to enforce RSM liveness 
in all current implementations of \trustBFT{} protocols. 
A quorum of $\f+1$ replicas only guarantees that one honest replica will commit, 
execute $\Transaction{}$, and reply to the client with the transaction result. 
All other replicas may act byzantine and fail to reply. 
In this situation, the sole honest replica 
replies to the client. 
}

\revise{
Existing \trustBFT{} protocols can be modified to support \RSM{} liveness, but at additional cost. 
This cost is often higher than the $3\f+1$ setup that they sought to improve on. 
There are three ways to address this issue: checkpointing, added latency, broadcasting.

\begin{enumerate}[nosep,wide]
\item {\em Checkpointing.} 
Periodic checkpoints will eventually bring honest replicas up-to-date and disseminate the necessary commit certificates. 
Unfortunately, this implies that clients will incur latency that is directly dependent on the checkpoint frequency 
(which tend to be relatively infrequent). 
Moreover, checkpoints require only $\f+1$ replicas to participate, and thus may not immediately include the necessary honest replica. 

\item {\em Added Latency.} 
A replica could, upon executing the transaction, include both the output and the commit certificate when replying to the client. 
The client could then, after a timeout, broadcast the commit certificates to all other replicas, 
thus informing them that the transaction is committed and can safely be executed.  
This approach introduces an additional round-trip (from $3$ to $4$ for \pbftea{}), 
at a time when consensus protocols are concerned about latency~\cite{minbft,basil}.  
As clients may not be located near the \RSM{} replicas, the added latency may be significant. 
Notice that the byzantine clients may fail to forward the commit certificate, further delaying the processing of subsequent (honest) clients.

\item {\em Broadcast.} Alternatively, upon committing a transaction $\Transaction{}$, 
replicas could systematically and preemptively broadcast the commit-certificate for $\Transaction{}$ 
to other replicas in the system. 
This additional all-to-all communication phase may cause a significant throughput drop~\cite{hotstuff}, 
especially when $\f$ is large.
\end{enumerate}

{\bf What about $3\f+1$?}  
Moving to $3\f+1$ and quorums of $2\f+1$ ensures that all committed operations will be committed at $\f+1$ honest replicas,
thus guaranteeing that the client will receive $\f+1$ responses for all operations.
}

%% file: unsafe.tex
\revise{
\section{Safety under Rollbacks}
}
\label{s:unsafe}

Existing \trustBFT{} protocols require some state to be {\em persisted} on the trusted hardware, corresponding to the logged requests or the counter values.
These systems rely on this property to {\em guarantee} safety. 
Despite any failures or attacks, these protocols expect this state to remain uncorrupted and available.
Unfortunately, realizing this assumption in available implementations of trusted hardware is challenging. 
Intel SGX enclaves, for instance, are the most popular
platform for trusted computing~\cite{scone,ccf,graphene-sgx,enclavedb}, but can suffer from
power-failures and rollback attacks. 
While recent works try to mitigate these attacks, they remain limited in scope
and have high costs~\cite{sgx-survey}. 
Unfortunately, hardware that has been shown to resist these attacks, such as SGX persistent counter~\cite{sgx} or TPM~\cite{tpm}, is
prohibitively slow: they have upwards of tens of milliseconds access latency and support only a limited number of writes~\cite{adamcs,sgx,memoir}.

Persistent state, is, as we show below, necessary for correctness
of \trustBFT{} protocols; a rollback attack can cause a node to equivocate. 
Once equivocation is again possible in a \trustBFT{} protocol, a single malicious
node can cause a {\em safety violation}.
To illustrate, we consider the following run of the \revise{\MinBFT{}} protocol. 
Let $\Faulty{}$ be the set of $\f$ Byzantine replicas, $D$ and $G$ a set of respectively 
$\f$ and $1$ honest replicas.

Assume that the primary $\Primary{}$ is byzantine and 
all replicas in $\Faulty$ intentionally fail to send replicas in $D$ any messages.
As is possible in a partially synchronous system, we further assume that the 
\revise{$\PName{Prepare}$ messages} from the replica in $G$ 
to those in $D$ are delayed.
$\Primary{}$ asks its trusted component to generate an attestation for a transaction 
$\Transaction{}$ to be ordered at sequence number $1$.
Following this, $\Primary{}$ sends a $\MName{Preprepare}$ message for $\Transaction$ 
to all replicas in $\Faulty$ and $G$. 
These $\f+1$ replicas are able to \revise{prepare $\Transaction$}, 
and they execute $\Transaction$ and reply to the client.
As a result, the client receives $\f+1$ identical responses and 
marks $\Transaction$ complete. 
Now, assume that the byzantine primary $\Primary$ rollbacks 
the state of its trusted component $\Trust{\Primary{}}$.
Following this, $\Primary{}$ ask its $\Trust{\Primary{}}$ to generate an attestation for a transaction 
$\Transaction{'}$ to be ordered at sequence number $1$.
Next, $\Primary{}$ sends a $\MName{Preprepare}$ message for $\Transaction{'}$ 
to all replicas in $D$.
Similarly, these replicas will be able to prepare and execute $\Transaction{'}$ and 
the client will receive $\f+1$ responses.
There is a safety violation as replicas in $D$ and $G$ have executed two 
different transactions at the same sequence number.
 
\revise{
{\bf How can we solve this?}
The straightforward approach is to replace all vulnerable enclave accesses 
with TPMs or persistent counters. 
While this solution may become viable in the future, we highlight in Figure~\ref{fig:real-world} 
that it is still impractical. 

{\bf What about 3f+1?} 
We propose a set of protocols that, by increasing the replication factor to $3\f+1$, 
reduce this overhead significantly: 
they limit the use of TPMs to once per transaction ($\BigO{\n}$ times for current \trustBFT{} protocols).
}

%% file: parallel.tex
\section{\revise{Lack of Parallelism}}
\label{s:parallel}
\trustBFT{} protocols are inherently sequential: they order client
requests one at a time and cannot support parallel consensus invocations.
\revise{ 
Pipelining consensus phases can mitigate the performance impact of this approach: 
it allows for the {\em Preprepare} phase of transaction $i+1$ to begin directly after the 
{\em Preprepare} phase of $i$ (similarly for {\em Prepare} and {\em Commit}). 
It does not address the root cause of the problem: 
the sequentiality of \trustBFT{} consensus protocols creates an artificial throughput bound 
on the throughput they can achieve (batch size / number phases $\times$ RTT). 
This is in direct contrast to traditional \BFT{} protocols which are parallel in nature: 
replicas can attempt to commit transaction $i+1$ concurrently with transaction $i$. 
}
As such, their throughput is bound by the available resources in the system.  
\revise{
Sequential consensus protocols perform especially poorly in the WAN-area as their throughput is directly proportional to phase latency.
}
\Hybster{}~\cite{hybster} attempts to mitigate this issue
by allowing each of the $\n$ replicas to act as a primary in parallel, 
but each associated consensus invocation is sequential.

\revise{
To illustrate why \trustBFT{} protocols cannot run 
consensus of two transactions in parallel, we assume the following run of \MinBFT{}.
}
Assume that the primary $\Primary{}$ allows consensus invocations of 
transactions $\Transaction_i$ and $\Transaction_j$ to proceed in parallel.
This implies that a replica $\Replica$ may receive $\MName{Preprepare}$ message for 
$\Transaction_j$ before $\Transaction_i$.
Further, on receiving a message, $\Replica{}$ calls the $\MName{Append}$ function to access its 
$\Trust{\Replica}$, which in turn attests this message for this specific order.

In our case, $\Replica$ would call $\MName{Append}$ on $\Transaction_j$ 
(before $\Transaction_i$) and will receive  an attestation 
$\SignMessage{\Message{Attest}{\tid,j,\Transaction_j}}{\Trust{\Replica}}$, 
which it will forward to all the replicas.
Now, when $\Replica{}$ receives $\MName{Preprepare}$ message for $\Transaction_i$, 
its attempt to call the $\MName{Append}$ function would fail. 
Its $\Trust{\Replica}$ ignores this 
message as $i < j$ and $\Trust{\Replica}$ cannot process a lower 
sequence number request.
The consensus for $\Transaction_i$ will not complete, stalling progress.

\revise{
{\bf What about $3\f+1$?}
By increasing the replication factor back to $3\f+1$, 
we can design a protocol that has higher throughput {\em per replica} than the 
$2\f+1$ approach (Figure~\ref{fig:eval-plots}(ii)).
This is a consequence of reduced number of trusted hardware accesses, 
which permits parallel consensus invocations at each replica.
}

%% file: extend.tex
\section{\lftBFT{} Protocols}
\label{s:extend}
The previous sections highlighted several significant limitations with
existing \trustBFT{} approaches, all inherited from their lower replication factor.  
In this section, we make two claims: 
(i) $2\f+1$ is simply \textit{not enough}:
it either impacts responsiveness or requires an extra phase of all-to-all communication (\S\ref{s:live});
it requires the use of slow persistent trusted counters for every message (\S\ref{s:unsafe}); and 
it sequentializes consensus decisions(\S\ref{s:parallel}). 
(ii) Trusted components {\em are still beneficial} to \BFT{} consensus if used with $3\f+1$ replicas 
as they can be used to reduce either the number of phases or the communication complexity.
To this effect, we present \lftBFT{}, a new set of \BFT{} consensus protocols that make use of
trusted components. 
These protocols satisfy \revise{both the \RSM{} and consensus} safety/liveness conditions described in Section~\ref{s:trust}, 
and, through the use of $3\f+1$ replicas, achieve the following appealing performance properties:
\begin{enumerate}[wide,nosep,label=(G\arabic*),ref={G\arabic*}]
\item \label{g:ooo} \revise{{\bf Parallel Consensus.} 
\lftBFT{} protocols allow replicas to process consensus invocations concurrently.}
\item \label{g:onetc} {\bf Minimal Trusted Component Use.}
\lftBFT{} protocols require accessing a single trusted component per transaction rather
        than one per message.
\revise{
This is especially important when using TPMs or persistent counters as the counter/logging service to preclude rollback attacks.
}
\item \label{g:nolog} {\bf No Trusted Logging.}
Moreover, \lftBFT{} protocols maintain low memory utilization at trusted components as they do 
not require trusted logging.
\end{enumerate}

\subsection{Designing a \lftBFT{} protocol}

\par We make three modifications to \trustBFT{} protocols. Together, these steps are sufficient
to achieve significantly greater performance, and better reliability.

\par First, we modify the $\MName{Append}$ functionality. Recall that the participants
use this function to bind a specific message with a counter value.
The value of this counter can be supplied by the replica but must 
increase monotonically; no two messages can be bound to the same value. We restrict this
function to preclude replicas from supplying their own sequence number, and instead have
the trusted component increment counters internally, thus ensuring that counter values will remain
contiguous. 
\begin{description}
\item[$\Message{AppendF}{\tid,x}$] -- Assume the $\tid$-th counter 
of $\Trust{\Replica}$ has value $\tct$.
This function increments $\tct$ to $\tct+1$, 
associates $\tct$ with message $x$, and returns 
an attestation $\SignMessage{\Message{Attest}{\tid,\tct,x}}{\Trust{\Replica}}$
as a proof of this binding.
\end{description}
This change is necessary to support \revise{parallel} consensus instances efficiently: while multiple
transactions can be ordered in parallel, the execution of these transactions must still take place
in sequence number order.  A Byzantine replica could stall the system's progress by issuing a sequence number that is far in the future, 
causing honest replicas to trigger frequent view changes to "fill" the gap with no-ops~\cite{pbftj}.

\revise{
\begin{description}
\item[$\Message{Create}{\tct}$] --
Creates a new counter with identifier $\tid$ and initial counter value $\tct$, 
such that no previous counter has an identifier $\tid$. 
This function also returns an attestation $\SignMessage{\Message{Attest}{\tid,\tct}}{\Trust{\Replica}}$.
\end{description}
We additionally make use of the standard functionality of creating new counters~\cite{trinc,sgx}. 
This function helps the new primary (post view change) to re-start consensus on previously proposed 
requests. 
For each new counter that a replica creates, it has to share a certificate (attestation) that proves 
the newness of this counter. 
}

\par Second, we reduce the number of calls to trusted counters. 
The primary is the only replica to solicit a new value from the trusted counter. All other participants simply validate the
trusted counter's signature when receiving a message from the primary. Specifically, upon receiving a client request $m := \Transaction{}$, the primary 
invokes $\Message{AppendF}{\tid,m}$ to bind a unique counter value $\tct$ to $m$ and returns an attestation 
$\SignMessage{\Message{Attest}{\tid,\tct,m}}{\Trust{\Primary}}$. The primary then forwards this attestation as part of its first consensus phase.
Crucially, replicas, upon receiving this message, simply verify the validity of the signature but no longer access their trusted components.

\par Finally, we increase the quorum size necessary to proceed to the next phase of consensus to $2\f+1$. This higher quorum size guarantees
that any two quorums will intersect in at least $\f+1$ distinct replicas (and thus in one honest replica). 
Forcing an honest replica to be part
of every quorum makes the need for accessing a trusted counter redundant as this replica will, by definition, never equivocate.

\changebars{}{{\bf View Change.} 
\lftBFT{} protocols \changebars{rely on a similar view-change protocol as their \trustBFT{} counterparts}{that require an explicit view-change protocol to replace 
a byzantine primary have access to 
a similar view-change protocol as their \trustBFT{} counterparts}.
However, like \BFT{} protocols, a view-change only takes place in a \lftBFT{} 
protocol when at least $2\f+1$ replicas agree to do the same.
Further, when at least $\f+1$ replicas request the replacement of primary $\Primary{}$ 
by broadcasting $\MName{ViewChange}$ messages, other honest replicas agree to 
support this mutiny. 

A $\MName{ViewChange}$ message from a replica $\Replica{}$ includes, 
for each request it received since the last checkpoint:
(i) a valid $\MName{Preprepare}$ message with a sequence number from $\Trust{\Primary{}}$, and
(ii) $2\f+1$ $\MName{Prepare}$ messages for each request it has committed.

Once the next primary $\Primary{'}$ receives $\MName{ViewChange}$ messages from at least 
$2\f+1$ replicas, it creates a $\MName{NewView}$ message and broadcasts it to all the replicas. 
Further, the trusted component $\Trust{\Primary{'}}$ at $\Primary{'}$ 
is now active and its counter is initialized to the value of last committed request.

{\bf Checkpoints.}
\lftBFT{} protocols employ same checkpoint protocol as their \trustBFT{} 
counterparts with one modification:  each replica $\Replica{}$ marks a checkpoint when it 
receives identical $\MName{Checkpoint}$ messages from at least $2\f+1$ other replicas (rather than $f+1$).
Each $\MName{Checkpoint}$ message includes, for each request that $\Replica$ has committed since the last checkpoint, 
the corresponding $\MName{Preprepare}$ message and $2\f+1$ $\MName{Prepare}$ messages.

{\bf Recovery.} 
Any failed replica $\Replica{}$ can easily recover either through $\MName{Checkpoint}$ or $\MName{NewView}$ messages. 
To recover, $\Replica{}$ needs to receive evidence of each request from at least $\f+1$  replicas.
}

\subsection{Case Study: \minLFT}

\begin{figure}[t]
    \begin{myprotocol}
	\INITIAL{Initialization:}{\newline
	{%\color{colN}
	// $\tid$ be the latest counter with value $\tct$.\\
	// $v$ be the view-number, determined by the identifier of primary.
	}}
	\SPACE

	%% Client request
        \TITLE{Client-role}{used by client $\Client{}$ to process transaction $\Transaction$}
        \STATE Sends $\SignMessage{\Transaction{}}{\Client{}}$ to the primary $\Primary{}$.
        \STATE Awaits receipt of messages $\Message{\Name{Response}}{\SignMessage{\Transaction{}}{\Client{}}, k, v, r}$ from 
		$\f+1$ replicas.
        \STATE Considers $\Transaction{}$ executed, with result $r$, as the $k$-th transaction.\label{fig:alg:recv}
        \SPACE 

	%% Primary receives client request
        \TITLE{Primary-role}{running at the primary $\Primary{}$}
	 \EVENT{$\Primary{}$ receives $\SignMessage{\Transaction{}}{\Client{}}$}
		\STATE Calculate digest $\Digest \GETS \Hash{\SignMessage{\Transaction{}}{\Client{}}}$.
		\STATE $\{k, \Sig\} \GETS \Message{AppendF}{\tid, \Digest}$ \label{minbft:append}
		\STATE Broadcast $\Message{\Name{Preprepare}}{\SignMessage{\Transaction{}}{\Client{}},\Digest, k, v, \Sig}$.
	 \ENDEVENT
	 \SPACE

	\TITLE{Non-Primary Replica-role}{running at the replica $\Replica{}$}
	 %% Non-Primary receives Pre-Prepare message.
	 \EVENT{$\Replica{}$ receives $\Message{\Name{Preprepare}}{\SignMessage{\Transaction{}}{\Client{}},\Digest,k, v, \Sig}$ from $\Primary{}$ such that:
		\begin{itemize}[nosep]
		\item Message is well-formed and attestation $\Sig$ is valid.
		\item $\Replica{}$ did not accept a $k$-th proposal from $\Primary{}$.
		\end{itemize}
	 }
		\STATE Broadcast $\Message{\Name{Prepare}}{\Digest,k,v,\Sig}$.	
	 \ENDEVENT
	 \SPACE

	\TITLE{Replica-role}{running at {\bf any} replica $\Replica{}$}
	%% Replica receives Prepare message.
	 \EVENT{$\Replica{}$ receives $\Message{\Name{Prepare}}{\Digest,k,v,\Sig}$ messages from $2\f+1$ replicas such that:\\
		\qquad all the messages are well-formed and identical. 
 	 }
	 
		\IF{$\Replica$ has executed transaction with sequence number $k-1$ $\wedge$ $k > 0$} 
			\STATE Execute $\Transaction$ as the $k$-th transaction.
			\STATE Let $r$ be the result of execution of $\Transaction$ (if there is any result).
			\STATE Send $\Message{\Name{Response}}{\SignMessage{\Transaction{}}{\Client{}}, k,v, r}$ to $\Client$.\label{fig:pa:inform}
		\ELSE
			\STATE Place $\Transaction{}$ in queue for execution.
		\ENDIF
	 \ENDEVENT
	 \SPACE

	%% Trusted Component receives client request
        \TITLE{Trusted-role}{running at the trusted component $\Trust{\Primary{}}$}
	 \EVENT{$\Trust{\Primary{}}$ is accessed through $\Message{AppendF}{\tid, \Digest}$}
		\STATE Increment $\tid$-th counter; $\tct := \tct + 1$
		\STATE $\Sig \GETS$  Generate attestation $\SignMessage{\Message{Attest}{\tid,\tct,\Digest}}{\Trust{\Primary}}$.
		\STATE Return $\{\tct, \Sig\}$.
	 \ENDEVENT

    \end{myprotocol}
    \caption{\minLFT{} protocol (failure-free path).}
    \label{alg:minlft}
\end{figure}

We apply our transformations to \MinBFT{}~\cite{minbft}, a two-phase \trustBFT{} protocol that makes use of trusted counters. 
\MinBFT{} requires one less phase than \pbft{} and \pbftea{} (two-phases total): as the primary
cannot equivocate, it is safe to commit a transaction in \MinBFT{} after receiving $\f+1$ $\MName{Prepare}$ messages.  
\minLFT{}, the {\em new protocol} that we develop, preserves this property, but remains responsive and makes minimal use of trusted
components (once per consensus). 
The view-change and checkpointing protocols
remain identical to the \pbft{} view-change, we do not discuss them in detail here. 
%Similarly, we omit view numbers fromcour description. 
We include pseudocode in Figure~\ref{alg:minlft}.

As stated, \minLFT{} consists of two phases. 
Upon receiving a transaction $\Transaction$, the primary $\Primary{}$ of view $v$ requests its
trusted component to generate an attestation for $\Transaction$.
This attestation $\SignMessage{\Message{Attest}{\tid,\tct,m}}{\Trust{\Primary{}}}$ 
states that $T$ will be ordered at position $\tct$ (Line~\ref{minbft:append}, Figure~\ref{alg:minlft}). The primary broadcasts
a $\MName{Preprepare}$ with this proof to all replicas. 
When a replica $\Replica{}$ receives a valid $\MName{Preprepare}$ 
message in view $v$, it marks the transaction $\Transaction$ as \textit{prepared}. 
Prepared transactions have the property that no
conflicting transaction has been prepared for the same counter value in the same view. 
In the \pbft{} protocol,
an additional round is necessary to mark messages as prepared, as replicas can equivocate. 
The replica $\Replica{}$ then broadcasts a $\MName{Prepare}$ message in support of $\Transaction$ and includes the attestation. 
When $\Replica$ receives $\MName{Prepare}$ messages from $2\f+1$ distinct replicas in the same view,
it marks $\Transaction$ as {\em committed}.  
$\Replica$ will execute $\Transaction$ once all transactions with sequence numbers smaller than $\tct$
have been executed. The client marks $\Transaction$ as complete when it receives matching responses from $\f+1$ replicas.

Replicas initiate the view-change protocol when they suspect that the primary has failed. 
As stated previously,
the view-change logic is identical to the \pbft{} view-change; we only describe it here briefly. 
A replica in view $v$ enters the view change by broadcasting a $\MName{ViewChange}$ message to all replicas. 
Each $\MName{ViewChange}$ message sent by a replica $\Replica$ 
includes all the valid \textit{prepared} and \textit{committed} messages received by $\Replica$ with relevant proof 
(the trusted component attestation for the $\MName{Preprepare}$ message and the $2\f+1$ $\MName{Prepare}$ messages for committed messages).
The new primary starts the new view if it receives $\MName{ViewChange}$ message from $2\f+1$ replicas for view $v+1$.

\subsection{Case Study: \minLFTZ{}} 
We now transform \MinZZ~\cite{minbft} into another novel \lftBFT{} protocol, \minLFTZ{}. 
\MinZZ{} is a \trustBFT{} protocol, which follows the design proposed by \ZZ~\cite{zyzzyva}. 
\ZZ{} introduces a \BFT{} consensus with a single-phase fast-path (when all replicas are honest and respond) and a two-phase slow-path. 
\MinZZ{} uses trusted counters to reduce the replication factor from $3\f+1$ to $2\f+1$. 
The cost of transforming \MinZZ{} to \minLFTZ{} is that, once again, we use $3\f+1$ replicas. 
However, there are several benefits (1) \minLFTZ{} can always go fast-path as 
it only requires $\n-\f$ matching responses (compared to the $\n$ for both \MinZZ{} and \ZZ{}). 
This helps improve performance under byzantine attacks, which past work has demonstrated is an issue for \ZZ~\cite{upright,aardvark,geobft}. 
(2) \minLFTZ{} minimizes the use of trusted components: a single access to a
trusted counter is required at the primary per consensus invocation.
(3) \minLFTZ{}'s view-change is significantly simpler than \ZZ's view-change. 
View-change protocols are notorious complex and error-prone~\cite{sbft,zyzzyva-unsafe} to design and implement; 
a simple view-change protocol thus
increases confidence in future correct deployments and implementations.
%(4) \minLFTZ{} does not face safety bug as \ZZ{}~\cite{zyzzyva-unsafe}.
We present pseudocode in Figure~\ref{alg:minlftz}. 

\begin{figure}[t]
    \begin{myprotocol}
	\INITIAL{Initialization: }{
	{%\color{colN}
	// $\tid$ be the latest counter with value $\tct$.
	}}
	\SPACE

	%% Client request
        \TITLE{Client-role}{used by client $\Client{}$ to process transaction $\Transaction$}
        \STATE Sends $\SignMessage{\Transaction{}}{\Client{}}$ to the primary $\Primary{}$.
        \STATE Awaits receipt of messages $\Message{\Name{Response}}{\SignMessage{\Transaction{}}{\Client{}}, k, v, r}$ from 
		$2\f+1$ replicas.
        \STATE Considers $\Transaction{}$ executed, with result $r$, as the $k$-th transaction.\label{fig:zz:recv}
        \SPACE 

	%% Primary receives client request
        \TITLE{Primary-role}{running at the primary $\Primary{}$ of view $v$}
	 \EVENT{$\Primary{}$ receives $\SignMessage{\Transaction{}}{\Client{}}$}
		\STATE Calculate digest $\Digest \GETS \Hash{\SignMessage{\Transaction{}}{\Client{}}}$.
		\STATE $\{k, \Sig\} \GETS \Message{AppendF}{\tid, \Digest}$
		\STATE Broadcast $m \GETS \Message{\Name{Preprepare}}{\SignMessage{\Transaction{}}{\Client{}},\Digest,k,v, \Sig}$.
		\STATE Execute($m$).
	 \ENDEVENT
	 \SPACE

	\TITLE{Non-Primary Replica-role}{running at the replica $\Replica{}$}
	 %% Non-Primary receives Pre-Prepare message.
	 \EVENT{$\Replica{}$ receives $m \GETS \Message{\Name{Preprepare}}{\SignMessage{\Transaction{}}{\Client{}},\Digest,k,v, \Sig}$ from $\Primary{}$ such that:
		\begin{itemize}[nosep]
		\item Message is well-formed and attestation $\Sig$ is valid.
		\item $\Replica{}$ did not accept a $k$-th proposal from $\Primary{}$.
		\end{itemize}
	 }
		\STATE Execute($m$).	
	 \ENDEVENT
	 \SPACE

	%% Trusted Component receives client request
        \TITLE{Trusted-role}{running at the trusted component $\Trust{\Primary{}}$}
	 \EVENT{$\Trust{\Primary{}}$ is accessed through $\Message{AppendF}{\tid, \Digest}$}
		\STATE Increment $\tid$-th counter; $\tct := \tct + 1$
		\STATE $\Sig \GETS$  Generate attestation $\SignMessage{\Message{Attest}{\tid,\tct,\Digest}}{\Trust{\Primary}}$.
		\STATE Return $\{\tct, \Sig\}$.
	 \ENDEVENT
	 \SPACE

	 \FUNC{Execute}{message: $m$}
		\IF{$\Replica$ has executed transaction with sequence number $k-1$ $\wedge$ $k > 0$} 
			\STATE Execute $\Transaction$ as the $k$-th transaction.
			\STATE Let $r$ be the result of execution of $\Transaction$ (if there is any result).
			\STATE Send $\Message{\Name{Response}}{\SignMessage{\Transaction{}}{\Client{}}, k, v, r}$ to $\Client$.\label{fig:zz:inform}
		\ELSE
			\STATE Place $\Transaction{}$ in queue for execution.
		\ENDIF
	 \ENDFUNC

    \end{myprotocol}
    \caption{\minLFTZ{} protocol (common-case).}
    \label{alg:minlftz}
\end{figure}

\par \textbf{Common Case.} 
A client $\Client$ submits a new transaction $\Transaction{}$ by sending a signed message  $\SignMessage{\Transaction}{\Client{}}$ to the primary $\Primary{}$.
The primary $\Primary{}$ invokes the $\Message{AppendF}{\tid,m}$ function,
binding the transaction to a specific counter value $\tct$ and returning an attestation $\SignMessage{\Message{Attest}{\tid,\tct,m}}{\Trust{\Primary}}$
as proof. 
This step prevents $\Primary$ from assigning
the same sequence number $\tct$ to two conflicting messages $m$ and $m'$.
The primary then forwards this attestation along with the transaction to 
all replicas. 
Replicas, upon receiving this message, execute the transaction in sequence order, and reply directly to the client with the response. 
The client marks the transaction $\Transaction{}$ complete when it receives $2\f+1$ identical responses in matching views.

{\bf View Change.} 
If the client does not receive $2\f+1$ matching responses, it
re-broadcasts its transaction to all replicas; the primary may have been malicious and failed to forward its request. Upon receiving this broadcast
request, a replica either (1) directly sends a response (if it has already
executed the transaction $\Transaction{}$) or, 
(2) forwards the request to the primary and starts a timer. 
If the timer expires before the replica receives a $\MName{Preprepare}$ message for $\Transaction{}$, it initiates a view-change. 
Specifically, the replica
enters view $v+1$, stops accepting any messages from view $v$, and
broadcasts a $\MName{ViewChange}$ message to all replicas. 
$\MName{ViewChange}$ messages include all requests for which $\Replica{}$ has received a $\MName{Preprepare}$ message. 

Upon receiving $\MName{ViewChange}$ messages from $2\f+1$ replicas in view $v+1$, 
the replica designated as the primary for view $v+1$ (say $\Primary'$) creates a $\MName{NewView}$ message and broadcasts it to all replicas. 
This message includes: 
(1) the set of $\MName{ViewChange}$ messages received by the primary as evidence, and 
(2) the (sorted-by-sequence-number) list of transactions that may have committed. 
The primary $\Primary'$ creates a new trusted counter
and sets it to the transaction with the lowest sequence number. 
$\Primary'$ then re-proposes all transactions in this list,
proposing specific no-op operations when there is a gap in the log between two re-proposed transactions. 

To re-propose these transactions,
the primary proceeds in the standard fashion: it accesses its trusted counter, obtains a unique counter value (setting
the counter to the transaction with the lowest sequence number ensures that sequence numbers remain the same across views), and
broadcasts the transaction and its attestation as part of a new $\MName{Preprepare}$ message. 
This mechanism guarantees that all transactions that \textit{could} have been perceived by the client as committed 
(the client receiving $2\f+1$ matching replies) will be re-proposed in the same order:
for the client to commit an operation, it must receive $2\f+1$ matching votes; one of those 
those votes is thus guaranteed to appear the \MName{NewView} message. 
Transactions that were executed by fewer than $2\f+1$ replicas, on the other hand, may \textit{not} be included in the new view, 
which may force some replicas to rollback.

%% file: proofs.tex
\revise{
\subsection{Proof Intuition}
We now state the theorems that prove correctness of our \lftBFT{} protocols.
For brevity, we only provide proof intuitions and refer the 
reader to Appendix~\ref{app:flexiBFT} for proofs.

\begin{theorem}\label{prop:non_divergent}
Let $\Replica_i$, $i \in \{1, 2\}$,  be two honest replicas that executed $\SignMessage{\Transaction_i}{\Client_i}$ as the $k$-th transaction of a given view $v$. 
If $\n > 3\f$, then $\SignMessage{\Transaction_1}{\Client_1} = \SignMessage{\Transaction_2}{\Client_2}$.
\end{theorem}

\begin{theorem}\label{theo3}
In a system $\Service = \{\Replicas, \Clients\}$ where 
$\abs{\Replicas} = \n = 3\f+1$, \minLFT{} and \minLFTZ{} protocols
guarantee consensus. 
\end{theorem}

To prove Theorem~\ref{theo3}, 
we first use Theorem~\ref{prop:non_divergent} to show that within a single view, 
both \minLFT{} and \minLFTZ{} guarantee consensus.
Next, we prove how a committed request persists across views. 
This is easy to prove for \minLFT{} as each replica only commits a transaction $\Transaction$
after it receives \MName{Prepare} messages for $\Transaction$ from a quorum of $2\f+1$ replicas.
However, in \minLFTZ{}, replicas can speculatively execute transactions.
So $\Transaction$ is marked committed if it is executed by $2\f+1$ replicas 
at the same sequence number, within a view.
This is safe as, in any following view, there will be at least one honest replica 
 that executed $\Transaction$.
}

%% file: eval.tex
\section{Evaluation}
\label{s:eval}
\revise{
The goal of our evaluation is to gauge how our \lftBFT{} protocols fare against their
\trustBFT{} and \BFT{} counterparts.
To do so, we ask three {\em core} questions. 
(1) How do our \lftBFT{} protocol perform and scale? (\S\ref{ss:observe} to \S\ref{ss:geo})
(2) What is the impact of failures? (\S\ref{ss:rep-fail})
(3) How will these protocols behave as hardware technology evolves? (\S\ref{ss:limit} and \S\ref{ss:hardware-cost})
}

\subsection{Implementation}
We use the open-source \ResilientDB{} fabric to implement all the consensus protocols~\cite{poe,rcc,geobft,kauri}. 
\ResilientDB{} supports all standard \BFT{} optimizations, including multithreading at each replica and both client
and server batching. 
The system relies on CMAC for \MAC{}, ED25519 for \DS{}-based signatures and SHA-256 for hashing.

{\bf SGX Enclaves.}
We use {\em Intel SGX for Linux}~\cite{sgx} to implement the abstraction of a trusted component at each replica. 
Specifically, we implement 
multiple monotonically increasing counters inside each enclave, which can be concurrently accessed by multiple threads through the
function {\tt GetSequenceNo(<counter-id>)}. 
This API call returns 
an attestation that includes the latest value of the specific counter and a \DS{} that proves that this counter value 
was generated by the relevant trusted component.
To highlight the potential of trusted components under the $3\f+1$ regime, we implement counters inside of the SGX enclave 
instead of leveraging Intel SGX Platform Services for trusted counters as they have prohibitively high latency and
are not available on most cloud providers. We highlight the trade-offs associated with the choice of trusted hardware in Section~\ref{ss:limit}.

\subsection{Evaluation Setup}

We compare our \lftBFT{} protocols against \revise{{\em eight}} baselines:
(i) \pbft{}~\cite{pbftj}, available with \ResilientDB{}, as it outperforms the BFTSmart's~\cite{bftsmart} 
implementation, which is single-threaded and sequential~\cite{poe,geobft,basil,kauri};
(ii) \ZZ~\cite{zyzzyva}, a linear single phase \BFT{} protocol where client expects responses from all the $3\f+1$ replicas;
(iii) \pbftea{}~\cite{a2m}, a three phase \trustBFT{} protocol; 
(iv) \MinBFT{}~\cite{minbft}, a two phase \trustBFT{} protocol; 
(v) \MinZZ{}~\cite{minbft}, a linear single phase \trustBFT{} protocol where client expects responses from all the $2\f+1$ replicas; and
(vi) \trustBFTc{}, a variation of \pbftea{} \changebars{we develop}{created by us} that supports parallel consensus invocations.
\revise{
(vii) o\minLFT{} and (viii) o\minLFTZ{} , variations of \minLFT{} and \minLFTZ{} with no parallel consensus invocations.
}
We do not compare against streamlined \BFT{} protocols such as Hotstuff~\cite{hotstuff} or \Damsys~\cite{damysus}, as their chained nature precludes concurrent 
consensus invocations.

We use the Oracle Cloud Infrastructure (OCI) and deploy up to $97$ replicas on
VM.Standard.E3.Flex machines (16 cores and \SI{16}{GiB} RAM) with \SI{10}{GiB} NICs.
Unless explicitly stated, we use the following setup: each system supports up to $\f=8$ Byzantine replicas. 
We intentionally choose a higher $\f$ to maximize the potential cost of increasing the replication factor to $3\f+1$. Clients run in
a closed-loop; each experiment runs for $180$ seconds (60 seconds warmup/cooldown) and we report average throughput/latency over three runs.
We adopt the popular {\em Yahoo Cloud Serving Benchmark} (YCSB), 
\changebars{}{which has been used to analyze several existing distributed applications}~\cite{ycsb,blockbench,geobft,deneva}.
YCSB \changebars{generates key-value store operations that}{allows us to generate key-value transactions that} access a database of
$\SI{600}{\kilo\nothing}$ records.

\changebars{}{The sizes of various messages communicated during any protocol are:
$\Name{Preprepare}$  (\SI{5392}{B}),
$\Name{Prepare}$ (\SI{216}{B}),
$\Name{Commit}$ (\SI{220}{B}), and
$\Name{Response}$ (\SI{2270}{B}).
}

\begin{figure}[t]
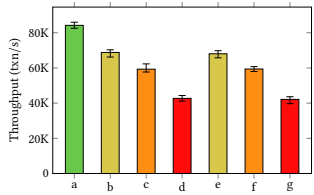

    \centering
    \graphPBFT
    \caption{\small Impact of trusted counter (TC) and signature attestations (SA) on \pbft{}. 
	[a] Standard \pbft{} protocol.
	[b] Primary (P) requires TC in Preprepare phase (Prep).
	[c] P requires TC and SA in Prep.
	[d] P requires TC and SA in all three phases.
	[e] All replicas require TC in Prep.
	[f] All replicas require TC and SA in Prep.
	[g] All replicas require TC and SA in all three phases.
   }
    \label{fig:pbft-costs}
\end{figure}

\begin{enumerate}[wide,nosep,label={\bf \ref{s:eval}.\arabic*},ref={\ref{s:eval}.\arabic*}]
\setcounter{enumi}{2}

\item {\bf Trusted Counter Costs.}
\label{ss:cost}
In Figure~\ref{fig:pbft-costs}, we quantify the costs of accessing trusted counters.
To do so, we run a \pbft{} implementation with a single worker-thread. 
Bar [a] represent our baseline implementation; we report
peak throughput numbers for each setup. 
Throughput degradation occurs when the primary replica needs to access the trusted component (Bar [b]). 
This degradation accelerates when the primary replica requires trusted component to perform signature attestations (Bar [c]), 
and needs to perform these operations during each phase of consensus (Bar [d]).
The drop in throughput from [a] to [d] is nearly $2\times$. Bars [e] to [g] extend the use of trusted components to non-primary replicas. 
        \changebars{The system is already bottlenecked at the primary replica (it must process more messages than replicas); this change
        thus has no impact on performance}{However, the relative throughput of the protocol remains the same as the protocol gets bottlenecked at the primary replica
        due to the higher number of messages that it must process.}

\begin{figure*}
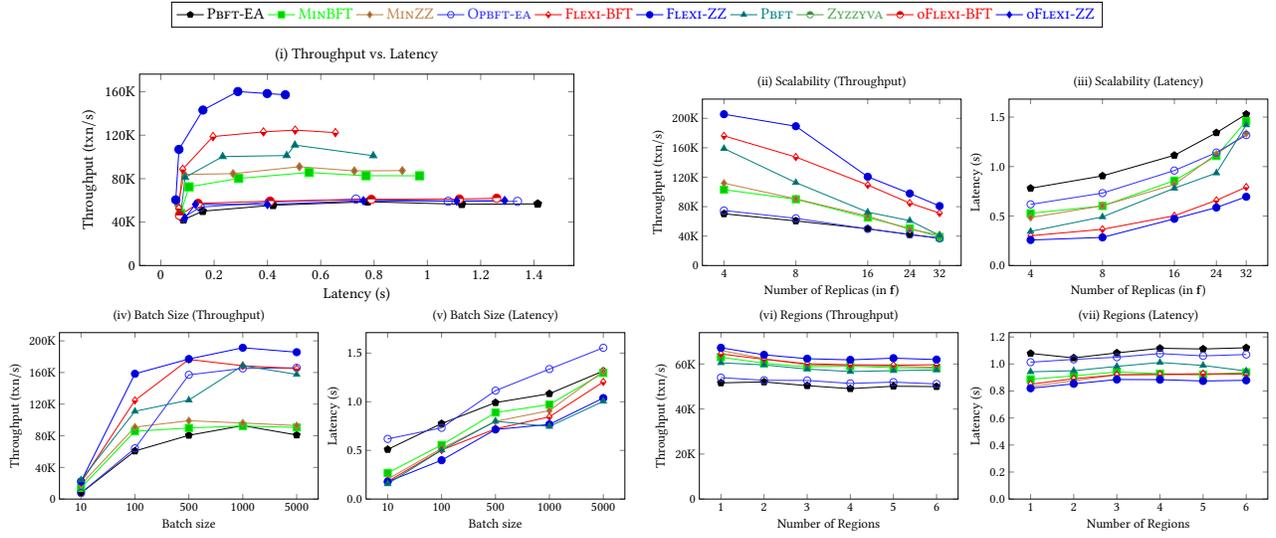

	%\vspace{-5mm}
    \centering
    \setlength{\tabcolsep}{1pt}
    \scalebox{0.6}{\ref{mainlegend}}\\[5pt]
    \begin{tabular}{cc@{\quad}cc}
    \multicolumn{2}{c}{\graphC} & \graphATP & \graphALat \\
    \graphBTP & \graphBLat & \graphDTP & \graphDLat

    \end{tabular}
    \caption{Throughput results as a function of number of clients, number of replicas, batch size and cross-site latency.
    We set the number of replicas in $\f$ as for \BFT{} and \lftBFT{} protocols 
    $\n = 3\f+1$ and for \trustBFT{} and \trustBFTc{} $\n = 2\f+1$.
    }
    \label{fig:eval-plots}
\end{figure*}

\item {\bf Throughput Results.}
\label{ss:observe}
        In Figure~\ref{fig:eval-plots}(i), we increase the number of clients \changebars{from \SI{4}{k} to \SI{80}{k}
and report on latency and throughput.}{ 
as follows: 
\SI{4}{k}, \SI{8}{k}, \SI{24}{k}, \SI{48}{k}, \SI{64}{k}, and \SI{80}{k} and 
        plot the throughput versus latency metric for all protocols}. \pbftea{} achieves 
the lowest throughput as it requires three phases for consensus and disallows parallel
consensus invocations. The reduced replication factor of $2\f+1$ does not help performance as
threads are already under-saturated:  the system is latency-bound rather than 
compute bound due to the protocol's sequential processing of consensus invocations.
\trustBFTc{} protocol attains up to $6\%$ higher throughput (and lower latency) than \pbftea{} as it supports parallel consensus invocations 
        but bottlenecks on trusted counter accesses at replicas\changebars{}{, as Figure~\ref{fig:pbft-costs} suggests}
Specifically, the replica's worker thread has to sign the outgoing message, and 
perform two verifications on each received message: 
(i) \MAC{} of the received message, and 
(ii) \DS{} of the attestation from the trusted component.
\MinBFT{} and \MinZZ{} achieve up to $47\%$ and $68\%$ higher throughputs than \pbftea{} respectively, as they reduce the number of phases
necessary to commit an operation (from three to two for \MinBFT, and from three to one in the failure-free case for \MinZZ).
Interestingly, \pbft{} yields better throughput than all \trustBFT{} protocols. 
The combination of parallel consensus invocations
and lack of overhead stemming from the use of trusted counters drives this surprising result. 
Our \minLFT{} and \minLFTZ{} protocols instead achieve up to $22\%$ and $58\%$ higher throughput than \pbft{},
(and up to $87\%$ and $77\%$ higher throughput over \MinBFT{} and \MinZZ{}). 
This performance improvement stems from reducing the number of phases,
accessing a trusted counter once per transaction, and 
permitting parallel consensus invocations.
\revise{
Note that supporting parallel consensus is key to these performance gains. 
Without this parallelism, our \lftBFT{} protocols perform worse than their
\trustBFT{} counterparts (o\minLFT{} yields $33\%$ less throughput than \MinZZ{}) 
as the primary needs to sequentially attest an additional $f$ messages.
}

\begin{figure}
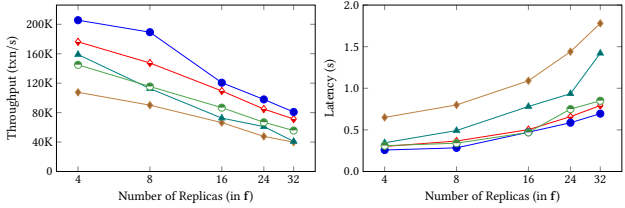

    \centering
    \setlength{\tabcolsep}{1pt}
    \begin{tabular}{c c}
    \graphETP & \graphELat
    \end{tabular}
    \caption{Impact of failure of one non-primary replica.}
    \label{fig:fail-plots}
\end{figure}

\item {\bf Scalability.}
\label{ss:scale}
Figures~\ref{fig:eval-plots}(ii) and~\ref{fig:eval-plots}(iii) summarize the protocols' scaling behavior as
the number of replicas increases from $\f=4$ to $\f=32$.
As expected, an increased replication factor leads to a proportional increase in the number of messages 
that are propagated and verified. This increased cost
leads to a significant drop in the latency and throughput of all protocols:
going from $\f=4$ to $\f=8$  causes \pbft{}, \minLFT{}, and \minLFTZ{}'s performance to drop $3.89\times$, $2.48\times$, and $2.54\times$
respectively.  \MinBFT{}, and \MinZZ{}'s throughput also drops by a factor of $2.66\times$ and $2.67\times$. 
This performance drop is larger for \BFT{} and \lftBFT{} protocols than for \trustBFT{} protocols as replicas are never fully saturated 
due to sequential consensus invocations.

\item {\bf Batching.}
\label{ss:batch}
Figures~\ref{fig:eval-plots}(iv) and~\ref{fig:eval-plots}(v) quantify the impact of batching
client requests as we increase the batch size from $10$ to \SI{5}{k}. As expected, the throughput of all protocols
increases as batch sizes increase \changebars{until communication becomes a bottleneck}{until the message size becomes too large, after which there are additional communication costs.}
\item {\bf Wide-area replication.}
\label{ss:geo}
For this experiment (Figures~\ref{fig:eval-plots}(vi) and~\ref{fig:eval-plots}(vii)), we distribute the replicas across five countries in six location: 
San Jose, Ashburn, Sydney, Sao Paulo, Montreal, and Marseille\changebars{, and use the regions in this order}{}.
We set $\f = 20$; $\n = 41$ and $\n = 61$
replicas for $2\f+1$ and $3\f+1$ protocols, respectively.%
\footnote{
As the LCM of numbers in the range $[1,6]$ is $60$, so we set $3\f+1 = 61$, which yields $\f=20$.}
\revise{
Latency and throughput remain mostly constant as the number of regions
increases. All evaluated systems require only quorums to make progress; these systems thus need to wait
only for responses of North American replicas (San Jose, Ashburn, Montreal). The increase in latency
or decrease in throughput is thus comparatively small.
}

\begin{figure}[t]
\vspace{-2mm}
\centering
\footnotesize
\begin{tabular}{!{\vrule width 1.5pt}c|c|c|c!{\vrule width 1.5pt}} \HLine{1.5pt}
%\rowcolor{black!15} & \multicolumn{3}{c!{\vrule width 1.5pt}}{Throughput (in \SI{}{txn/s})} \\  \cline{2-4}
\rowcolor{black!15}{Access cost (in \SI{}{ms})} & \minLFTZ{} & \MinZZ{} & \MinBFT{} \\ \hline
1.0	& \SI{87}{k} & \SI{49}{k}   & \SI{39}{k}     \\ \hline 
1.5	& \SI{67}{k} & \SI{48}{k}   & \SI{37}{k}     \\ \hline 
2.0	& \SI{50}{k} & \SI{47}{k}   & \SI{35}{k}     \\ \hline 
2.5	& \SI{40}{k} & \SI{40}{k}   & \SI{34}{k}     \\ \hline 
3.0	& \SI{34}{k} & \SI{33}{k}   & \SI{32}{k}     \\ \hline
10	& \SI{10}{k} & \SI{10}{k}   & \SI{10}{k}     \\ \hline
30	& \SI{3}{k}  & \SI{3}{k}    & \SI{3}{k}      \\ \hline
100	& $993$	     & $959$	    & $994$	         \\ \hline
200	& $494$	     & $479$	    & $496$	         \\ \HLine {1.5pt}
\end{tabular}
\caption{Peak throughput (in transactions per second) on varying the time taken to access a trusted counter 
while running consensus among $97$ replicas. 
}
\label{fig:real-world}
\end{figure}

\item {\bf Single Replica Failure.}
\label{ss:rep-fail}
Next, we consider the impact of failures on our protocols (Figure~\ref{fig:fail-plots}). 
Unlike \MinZZ{} and \ZZ{}, our \minLFTZ{} protocol's performance does 
not degrade as it can handle up to $\f$ non-primary replica failures on the fast path.
In contrast, both \MinZZ{} and \ZZ{} require their clients to receive responses from all replicas; 
in order to commit in a single round-trip.

\item {\bf Real-World Adoption}
\label{ss:limit}

This paper's objective is to highlight current limitations of existing
\trustBFT{} approaches, be it hardware-related or algorithmic. Trusted
hardware, however, is changing rapidly. Current SGX enclaves are subject to rollback attacks, but 
newer enclaves (Keystone, Nitro) may not be. Similarly, accessing current SGX persistent counters
or TPMs currently takes between \SI{80}{ms} to \SI{200}{ms} for TPMs and between
\SI{30}{ms} to \SI{187}{ms} for SGX~\cite{memoir,rote,trinc}. New technology is rapidly bringing this cost down;
counters like ADAM-CS~\cite{adam-cs} requires less than \SI{10}{ms}. 
Our final experiment aims to investigate
the current performance of \trustBFT{} protocols on both present and future trusted hardware. Our previous results
were obtained using counters inside of SGX enclave as hardware providing access to SGX persistent counters and TMPs
are not readily avalable on cloud providers. 
In this experiment, we gage the impact of throughput and latency as we increase
the time to access the trusted counter (Figure~\ref{fig:real-world}) on \minLFTZ{}, \revise{\MinBFT{} and \MinZZ{}} protocols. 
We run this experiment on $97$ replicas, and highlight that, for this setup \pbft{} yields \SI{40}{k txn/s}. 
We find that \minLFTZ{} 
\revise{outperforms all protocols as long as the latency} is less than \SI{2.5}{ms}. 
\revise{
Beyond this value, a single access to trusted hardware becomes the bottleneck; 
causing all protocols' performance to degrade to similar values 
(eg. at \SI{10}{ms}, \SI{10}{k} can be directly obtained by batch size $\times$ \SI{1}{s} $/$ \SI{10}{ms}).
}
This result highlights a path whereby, as trusted hardware matures,
\trustBFT{} protocols will be an appealing alternative to standard \BFT{} approaches.

\begin{figure}[t]
\centering
\footnotesize
\begin{tabular}{!{\vrule width 1.5pt}c|c|c!{\vrule width 1.5pt}} 	\HLine{1.5pt}
\rowcolor{black!15} Replicas (in $\f$) & \minLFTZ{} & \MinZZ{} 		\\ \hline
$4$ 	& $15813$	& $12431$     					\\ \hline 
$8$ 	& $7570$   	& $5329$     					\\ \hline 
$16$	& $2462$	& $2038$	         			\\ \hline
$24$	& $1341$	& $1002$					\\ \hline
$32$	& $834$		& $640$						\\ \HLine {1.5pt}
\end{tabular}
\caption{\revise{Throughput-per-machine: total system throughput/(number of replicas)}}
\label{fig:tput-per-hardware}
\end{figure}

\revise{
\item {\bf Throughput-Per-Machine.}
\label{ss:hardware-cost}
\trustBFT{} protocols seek to reduce the hardware necessary to deploy \BFT{} consensus; 
additional replicas increase operational complexity and resource costs. 
In fact, the high costs of accessing trusted hardware for every message combined with the lack of parallelism 
that results from a $2\f+1$ replication factor decreases overall system throughput. 
We find that reverting to $3\f+1$ actually {\em increases} 
the {\em throughput-per-machine} performance of the system (for the reasons outlined above). 
Per machine, $3\f+1$ \lftBFT{} protocol achieve higher throughput than a $2\f+1$ 
\trustBFT{} protocol (up to $30\%$, as shown in Figure~\ref{fig:tput-per-hardware}).
}

\end{enumerate}

%% file: related.tex
\section{Related Work}
\label{s:related}
There is long line of research on designing efficient \BFT{} protocols~\cite{pbftj,zyzzyva,poe,geobft,hotstuff,basil,rcc,kauri,next700bft,rbft}. 
As stated, \trustBFT{} protocols prevent replicas from equivocating, reducing the replication factor or the number of phases necessary to achieve safety. 
We summarized \pbftea{} in Section~\ref{s:trust} and now describe other \trustBFT{} protocols.

\CheapBFT{}~\cite{cheapbft} uses trusted counters and optimizes for the {\em failure-free} case by 
reducing the amount of {\em active} replication to $\f+1$. When a failure occurs, however,
\CheapBFT{} requires that all $2\f+1$ replicas participate in consensus. The protocol has the same
number of phases as \MinBFT{}; 
as higlighted in Sections~\ref{s:live} and~\ref{s:parallel}, the protocol is inherently sequential
and may not be responsive to clients.

\Hybster{}~\cite{hybster} is a meta-protocol that 
takes as an input an existing \trustBFT{} protocol 
(such as \pbftea{}, \MinBFT{}, etc.) and 
requires each of the $\n$ replicas to act as parallel primaries; a common deterministic execution framework will consume these local consensus logs
to execute transactions in order.
While multiple primaries improve concurrency, each primary locally invokes consensus in-sequence; each sub-log locally inherits the limitations
of existing \trustBFT{} protocols. There continues to be an artificial upper-bound on the amount of parallelism supported in the system. Moreover,
recent work shows that designing multiple primary 
protocols is hard as $\f$ of these primaries can be byzantine and can collude to 
prevent liveness~\cite{rcc,iss}.

{\em Streamlined protocols} like \hotstuffm{}~\cite{hotstuffm} and \Damsys~\cite{damysus}
follow the design of \hotstuff{}~\cite{hotstuff}. 
They linearize communication by splitting the all-to-all communication phases 
($\MName{Prepare}$ and $\MName{Commit}$) into two linear phases.
These systems additionally rotate the primary after each transaction, 
requiring $\f+1$ replicas to send their last committed message to 
the next primary. The next primary then selects the committed message with the highest view number as the 
baseline for proposing the next transaction for consensus.
\hotstuffm{} makes use of trusted logs; \Damsys{} requires its replicas to have trusted components that provide support 
for both logs and counters. 
Specifically, \Damsys{} requires two types of trusted components at each replica, an accumulator and a checker.
The primary leverages the accumulator to process incoming messages and 
create a certificate summarizing the last round of consensus. 
Each replica instead accesses the checker to generate sequence numbers using a monotonic counter and 
logs information about previously agreed transactions. These protocols once again suffer from a potential
lack of responsiveness; their streamlined nature precludes opportunities to support any concurrency~\cite{geobft,kauri}.

Microsoft's CCF framework uses Intel SGX to support building confidential and verifiable services~\cite{ccf,ia-ccf}. 
CCF provides a framework that helps to generate an audit trail for each request. 
To do so, they log each request and have it attested by the trusted components.
CCF provides flexibility of deploying any consensus protocol.

In the specific case of blockchain systems, Teechain~\cite{teechain} 
designs a two-layer payment network with the help of SGX.
Teechain designates trusted components as treasuries and only allows them to manage user funds. 
Teechain permits a subset of treasuries to be compromised, and 
it handles such attacks by requiring each fund to be managed by a group of treasuries.
Ekiden~\cite{ekiden} executes smart contracts directly in the trusted component for better
privacy and performance. 
\revise{Avoine et al.~\cite{graceful-fair-exchange} provide a good theoretical treatment of 
fair-exchange problem using trusted hardware.}

%% file: concl.tex
\section{Conclusion}
\label{s:concl}
In this paper, we identified three challenges with the design of existing 
\trustBFT{} protocols:
(i) they have limited responsiveness, 
(ii) they suffer from safety violations in the presence of rollback attacks, and
(iii) their sequential nature artificially bounds throughput. 
We argue that returning to $3\f+1$ is the key to fulfilling the potential of trusted components in \BFT{}.
Our suite of protocols, \lftBFT{}, supports parallel consensus instances, makes minimal use of trusted components,
and reduces the number of phases necessary to safely commit operations while also simplifying notoriously complex
mechanisms like view-changes. 
In our experiments, \lftBFT{} protocols outperform their \BFT{} and \trustBFT{} counterparts by $100\%$ and $185\%$ respectively.

%{, which, unlike their \BFT{} counterparts, require only 
%$2\f+1$ replicas; these replicas access their trusted components to agree on 
%an order for each client request.
%We show that these \trustBFT{} protocols:
%(i) face loss of responsiveness due to a single delayed message at a replica, 
%(ii) suffer from loss of safety due to rollback attacks on trusted components, and 
%(iii) enforce sequential consensus of client requests.
%To resolve these challenges, we present a novel suite of protocols, \lftBFT{} protocols, 
%which require one trusted component per consensus and permit concurrent consensus invocations.
%We design two \lftBFT{} protocols, and
%our results prove that our \lftBFT{} protocols outperform their \BFT{} and \trustBFT{} counterparts.
%}

%% file: flexibft.tex
\section{Proofs}
\label{app:flexiBFT}
We first prove that in \lftBFT{} protocols, no two honest replicas will execute two different requests at the same sequence number.

\begin{theorem}\label{prop:non_divergent}
Let $\Replica_i$, $i \in \{1, 2\}$,  be two honest replicas that executed $\SignMessage{\Transaction_i}{\Client_i}$ as the $k$-th transaction of a given view $v$. 
If $\n > 3\f$, then $\SignMessage{\Transaction_1}{\Client_1} = \SignMessage{\Transaction_2}{\Client_2}$.
\end{theorem}
\begin{proof}
Replica $\Replica_i$ only executed $\SignMessage{\Transaction_i}{\Client_i}$ after $\Replica_i$ received a well-formed $\MName{Preprepare}$ message 
($\Message{\Name{Preprepare}}{\SignMessage{\Transaction{}}{\Client{}},\Digest,k, v, \Sig}$) from the primary $\Primary$. 
This message includes an attestation $\Sig = \SignMessage{\Message{Attest}{\tid,\tct,\Digest}}{\Trust{\Primary}}$ from $\Trust{\Primary}$, which we assume 
cannot be compromised.
Let $S_i$ be the replicas that received $\MName{Preprepare}$ message for $\SignMessage{\Transaction_i}{\Client_i}$.
Let $X_i = S_i \difference \Faulty$ be the honest replicas in $S_i$. 
As $\abs{S_i} = 2\f+1$ and $\abs{\Faulty} = \f$, we have $\abs{X_i} = 2\f+1 - \f$. 
The honest replicas in $T_i$ will only execute the $k$-th transaction in view $v$ if it has a attestation from the trusted component at the primary. 
If $\SignMessage{\Transaction_1}{\Client_1} \neq \SignMessage{\Transaction_2}{\Client_2}$, then $X_1$ and $X_2$ must not overlap as the trusted 
component will never assign them the same sequence number.
Hence, $\abs{X_1 \union X_2} \geq 2(2\f+1 - \f)$. 
This simplifies to $\abs{X_1 \union X_2} \geq 2\f+2$, which contradicts $\n > 3\f$. 
Hence, we conclude $\SignMessage{\Transaction_1}{\Client_1} = \SignMessage{\Transaction_2}{\Client_2}$.
\end{proof}

Next, we show that \minLFT{} guarantees a safe consensus.

\begin{theorem}
In a system $\Service = \{\Replicas, \Clients\}$ where 
$\abs{\Replicas} = \n = 3\f+1$, \minLFT{} protocol
guarantees a safe consensus. 
\end{theorem}

\begin{proof}
If the primary $\Primary{}$ is honest, then from Theorem~\ref{prop:non_divergent}, we can conclude 
that no two replicas will execute different transactions for the same sequence number. 
This implies that all the honest replicas will execute the same transaction per sequence number.
If a transaction is executed by at least $\f+1$ honest replicas, then it will persist across views
as in any view-change quorum of $2\f+1$ replicas, there will be one honest replica that 
has executed this request and has a valid $\MName{Preprepare}$ message and $2\f+1$ $\MName{Prepare}$ 
messages corresponding to this request. 

If the primary $\Primary{}$ is byzantine, it can only prevent broadcasting the $\MName{Preprepare}$ 
messages to a subset of replicas. 
$\Primary{}$ cannot equivocate as it does not assign sequence numbers. 
For each transaction $\Transaction$, $\Primary{}$ needs to access its $\Trust{\Primary{}}$, which returns a 
sequence number $k$ and an attestation that $k$ binds $\Transaction{}$.
Further, Theorem~\ref{prop:non_divergent} proves that for a given view, no two honest 
replicas will execute different transactions for the same sequence number.
So, a byzantine $\Primary{}$ can send the $\MName{Preprepare}$ for $\Transaction{}$:
(i) to at least $2\f+1$ replicas, or
(ii) to less than $2\f+1$ replicas. 
In either cases, any replica $\Replica{}$ that receives $\Transaction{}$ will send the $\MName{Prepare}$ message. 
If $\Replica{}$ receives $2\f+1$ $\MName{Prepare}$ messages, it will execute $\Transaction{}$ and reply to the client.
Any remaining replica that did not receive $\Transaction{}$ will eventually timeout waiting for a request and 
trigger a $\MName{ViewChange}$.
If at least $\f+1$ replicas timeout, then a $\MName{ViewChange}$ will take place. 

If $\Transaction{}$ was prepared by at least $\f+1$ honest replicas, then this request will be part of the 
subsequent view. 
Otherwise, the subsequent view may or may not include $\Transaction$.
But this should not be an issue because such a transaction was not executed by any honest replica; 
no replica would have received $2\f+1$ $\MName{Prepare}$ messages.
Hence, system is safe even if this transaction is forgotten.

Each new view $v+1$ is led by a replica with identifier $i$, where $i = (v+1)~ \textrm{mod} ~\n$. 
The new primary waits for $\MName{ViewChange}$ messages from $2\f+1$ replicas, 
uses these messages to create a $\MName{NewView}$ message, and forwards these messages to all the replicas.
This $\MName{NewView}$ message includes a list of requests for each sequence number present in the $\MName{ViewChange}$ message.
The new primary needs to set its counter to the lowest sequence number of this list (may need to create a new counter).
Post sending the $\MName{NewView}$ message, the new primary re-proposes the $\MName{Preprepare}$ message for 
each request in the $\MName{NewView}$. 
Each replica on receiving the $\MName{NewView}$ message can verify its contents.
\end{proof}

Next, we show that \minLFTZ{} guarantees a safe consensus.

\begin{theorem}
In a system $\Service = \{\Replicas, \Clients\}$ where 
$\abs{\Replicas} = \n = 3\f+1$, \minLFTZ{} protocol guarantees 
a safe consensus.
\end{theorem}

\begin{proof}
If the primary $\Primary{}$ is honest, then from Theorem~\ref{prop:non_divergent}, we can conclude 
that no two replicas will execute different transactions for the same sequence number. 
This implies that all the honest replicas will execute the same transaction per sequence number.
If a transaction is executed by at least $\f+1$ honest replicas, then it will persist across views
as in any view-change quorum of $2\f+1$ replicas, there will be one honest replica that 
has executed this request and has a valid $\MName{Preprepare}$ message.

If the primary $\Primary{}$ is byzantine, it can only prevent broadcasting the $\MName{Preprepare}$ 
messages to a subset of replicas. 
$\Primary{}$ cannot equivocate as it does not assign sequence numbers. 
For each transaction $\Transaction$, $\Primary{}$ needs to access its $\Trust{\Primary{}}$, which returns a 
sequence number $k$ and an attestation that $k$ binds $\Transaction{}$.
Further, Theorem~\ref{prop:non_divergent} proves that for a given view, no two honest 
replicas will execute different transactions for the same sequence number.
So, a byzantine $\Primary{}$ can send the $\MName{Preprepare}$ for $\Transaction{}$:
(i) to at least $2\f+1$ replicas, or
(ii) to less than $2\f+1$ replicas. 
In either cases, any replica that receives $\Transaction{}$ will execute it and reply to the client, 
while the remaining replicas will eventually timeout waiting for a request and trigger a $\MName{ViewChange}$.
If at least $\f+1$ replicas timeout, then a $\MName{ViewChange}$ will take place. 
If $\Transaction{}$ was executed by at least $\f+1$ honest replicas, then this request will be part of the 
subsequent view. 
Otherwise, the subsequent view may or may not include $\Transaction$.
In such a case, any replica that executed $\Transaction$ would be required to rollback its state.
However, for any request if the client receives $2\f+1$ responses, it will persist across views 
because at least $\f+1$ honest replicas have executed that request.

Each new view $v+1$ is led by a replica with identifier $i$, where $i = (v+1)~ \textrm{mod} ~\n$. 
The new primary waits for $\MName{ViewChange}$ messages from $2\f+1$ replicas, 
uses these messages to create a $\MName{NewView}$ message, and forwards these messages to all the replicas.
This $\MName{NewView}$ message includes a list of requests for each sequence number present in the $\MName{ViewChange}$ message.
The new primary needs to set its counter to the lowest sequence number of this list (may need to create a new counter).
Post sending the $\MName{NewView}$ message, the new primary re-proposes the $\MName{Preprepare}$ message for 
each request in the $\MName{NewView}$. 
Each replica on receiving the $\MName{NewView}$ message can verify its contents.
\end{proof}